\documentclass[seceq,preprint]{ptptex}

\usepackage{amsmath}
\usepackage{amssymb}
\usepackage[dvipdfmx]{graphicx}

\preprintnumber{IU-TH-10
}

\title{%
Theory Including Future Not Excluded
}

\subtitle{Formulation of Complex Action Theory II }    

\author{%
Keiichi \textsc{Nagao}\footnote{E-mail: keiichi.nagao.phys@vc.ibaraki.ac.jp} 
and Holger Bech \textsc{Nielsen}\footnote{E-mail: hbech@nbi.dk}
}

\inst{
$^{*)}$Faculty of Education, Ibaraki University, Bunkyo 2-1-1, Mito 310-8512 \\
Japan \\
{\it and }\\
$^{**)}$
Niels Bohr Institute, University of Copenhagen, 
17 Blegdamsvej Copenhagen $\phi$, Denmark
}

\abst{

We study a complex action theory (CAT) 
whose path runs over not only past but also future.
We show that if we regard a matrix element defined in terms of the future state 
at time $T_B$ and the past state at time $T_A$ as an expectation value in the CAT, 
then we are allowed to have the Heisenberg equation, the Ehrenfest's theorem 
and the conserved probability current density. 
In addition we show that the expectation value at the present time $t$ 
of a future-included theory for large $T_B-t$ and large $t- T_A$ corresponds to 
that of a future-not-included theory with a proper inner product for large $t- T_A$.  
Hence the CAT with future explicitly present in the formalism 
and influencing in principle the past is not excluded phenomenologically, 
because the effects are argued to be very small in the present era. 
Furthermore we explicitly derive the Schr\"{o}dinger equation and Hamiltonian 
for the future state via path integral, and 
confirm that the Hamiltonian is given by the Hermitian conjugate of the Hamiltonian for the past state.
}

\begin{document}

\maketitle

\section{Introduction }

Quantum theories are properly formulated via the Feynman path integral (FPI). 
Usually the action $S$ is real, and it is thought to be more fundamental than 
the integrand $\exp(\frac{i}{\hbar}S)$. 
However, if we assume that the integrand is more fundamental
than the action in quantum theory, then it is naturally thought that
since the integrand is complex, the action could also be complex. 
Based on this speculation and other related works 
in some backward causation developments inspired by general relativity\cite{ownctl} 
and the non-locality explanation 
of fine-tuning problems \cite{nonlocal}, the complex action theory (CAT) 
has so far been studied intensively\cite{Bled2006,own}. 
The imaginary part of the action is thought to give some falsifiable predictions, 
and 
many interesting suggestions have been made for the Higgs mass\cite{Nielsen:2007mj}, 
quantum-mechanical philosophy\cite{newer1,Vaxjo2009,newer2}, 
some fine-tuning problems\cite{Nielsen2010qq,degenerate}, 
black holes\cite{Nielsen2009hq}, 
the de Broglie-Bohm particle, and a cut-off in loop diagrams\cite{Bled2010B}.

In refs.\cite{Bled2006,own,Nielsen:2007mj,newer1,Vaxjo2009, newer2,Nielsen2010qq,degenerate,Nielsen2009hq,Bled2010B} 
they studied a future-included version, i.e., 
the theory including not only a past time but also a future time 
as an integration interval of time. 
In contrast to these works, in refs.\cite{Nagao:2010xu, Nagao:2011za, Nagao:2011is}
we studied a future-not-included version. 
In ref.\cite{Nagao:2010xu} we analyzed 
the time development of some state by a non-Hermitian diagonalizable 
bounded Hamiltonian $H$, 
and showed that we can effectively obtain a Hermitian Hamiltonian after a long time 
development by introducing 
a proper inner product\footnote{A similar inner product was also studied 
in ref.\cite{Geyer}.} 
based on the speculation in ref.\cite{originsym}. 
If the Hermitian Hamiltonian is given in a local form, 
a conserved probability current density can be constructed with two kinds of wave functions. 
We note that the non-Hermitian Hamiltonian studied there is a generic one, 
so it does not belong to 
the class of PT symmetric non-Hermitian Hamiltonians, 
which has recently been intensively studied.\cite{PTsym_Hamiltonians,Geyer}

In ref.\cite{Nagao:2011za} 
introducing a philosophy to keep the analyticity 
in the parameter variables of FPI and defining a modified set of a complex conjugate, 
real and imaginary parts, Hermitian conjugates, and bras, 
we explicitly constructed non-Hermitian operators of coordinate and momentum, 
$\hat{q}_{new}$ and $\hat{p}_{new}$, and their eigenstates 
${}_m\langle_{new}~ q |$ and ${}_m\langle_{new}~ p |$ 
for complex $q$ and $p$ 
by utilizing coherent states of harmonic oscillators 
so that we can deal with complex $q$ and $p$. 
In addition, applying this complex coordinate formalism to the study of ref.\cite{Nagao:2010xu}, 
we showed that the mechanism for suppressing the anti-Hermitian part of the Hamiltonian 
after a long time development also works in the complex coordinate case.
In ref.\cite{Nagao:2011is} based on the complex coordinate formalism 
we explicitly examined the definitions of the momentum and Hamiltonian via FPI, 
and confirmed that they have the same forms as those in the real action theory (RAT).

Regarding other studies related to complex coordinates, 
in refs.~\cite{Garcia:1996np,Guralnik:2007rx} 
the complete set of solutions of the differential equations 
following from the Schwinger action principle has been obtained by generalizing 
the path integral to include sums over various inequivalent contours of integration in 
the complex plane. 
In ref.~\cite{Pehlevan:2007eq}, complex Langevin equations have been studied, and 
in refs.~\cite{Ferrante:2008yq}\cite{Ferrante:2009gk} a method to examine the complexified solution set has been investigated.

In this paper we study a future-included version of the CAT whose path runs over not only past 
but also future\cite{Bled2006} 
using both the complex coordinate formalism\cite{Nagao:2011za} 
and the mechanism 
for suppressing the anti-Hermitian part of the Hamiltonian\cite{Nagao:2010xu}. 
In ref.\cite{Bled2006}, one of the authors of this paper, H.B.N., and Ninomiya 
introduced not only the ordinary past state $| A (T_A) \rangle$ at the initial time $T_A$, but also 
a future state $| B (T_B) \rangle$ at the final time $T_B$, 
where $T_A$ and $T_B$ are set to be $-\infty$ and $\infty$ respectively. 
Here $| A (T_A) \rangle$ and $| B (T_B) \rangle$ 
time-develop according to the non-Hermitian Hamiltonian $\hat{H}$ and $\hat{H}_B$, respectively, 
where $\hat{H}_B$ is set to be equal to $\hat{H}^\dag$. 
They studied the matrix element of some operator ${\cal O}$ defined by 
\begin{equation}
\langle {\cal O} \rangle^{BA} 
\equiv\frac{ \langle B(t) |  {\cal O}  | A(t) \rangle }{ \langle B(t) | A(t) \rangle } , \label{expvalOBA0}
\end{equation}
where $t$ is the present time. 
In the RAT, such a future state as $|B \rangle$ 
is already introduced in ref.\cite{AAV} in a different context. 
The matrix element of eq.(\ref{expvalOBA0}), which is called the weak value, 
has also been intensively studied. 
For details of the weak value, see the reviews\cite{review_wv} and references therein. 
Eq.(\ref{expvalOBA0}) is a matrix element in the usual sense, but 
in a future-included version of the CAT 
we speculate that it can be regarded as the expectation value of ${\cal O}$ 
from the results that we obtain in this paper. 
As we shall see later, $\langle {\cal O} \rangle^{BA}$ 
allows us to have the Heisenberg equation. 
In addition, we shall confirm that it gives us Ehrenfest's theorem. 
Furthermore, we shall also see that a conserved probability current density can be constructed.  
Therefore we regard it as an expectation value in the future-included theory.

Here we note that since the future-included theory differs from ordinary quantum mechanics 
on two points -- the existence of the imaginary part of the action $S$ 
and that of the future state -- it seems excluded phenomenologically. 
So it is necessary that the future-included theory is not excluded, 
to show that usual physics is approximately obtained from it. 
Indeed, in ref.\cite{Bled2006}, 
an attempt was made to obtain a correspondence 
between the future-included theory and ordinary quantum mechanics, 
and it is speculated that
$\langle {\cal O} \rangle^{BA}$ becomes 
\begin{equation}
\langle {\cal O} \rangle^{AA} 
\equiv \frac{  \langle A(t) | {\cal O}  | A(t) \rangle }
{ \langle A(t) | A(t) \rangle } , \label{OAA}
\end{equation}
i.e. the expectation value of ${\cal O}$ in the future-not-included theory. 
We review this speculation and 
make it clear that there are points to be improved in the argument. 
Then we study $\langle {\cal O} \rangle^{BA}$ with more care concerning 
the inner product being obtained 
by using both the complex coordinate formalism and the mechanism for suppressing the anti-Hermitian 
part of the Hamiltonian, and show that $\langle {\cal O} \rangle^{BA}$ becomes 
an expectation value with a different inner product defined in the future-not-included theory. 
Next we show that the inner product can be interpreted as one of the possible proper 
inner products realized in the future-not-included theory. 
Thus we shall have the correspondence principle: 
the future-included theory for large $T_B-t$ and large $t-T_A$ is almost 
equivalent to the future-not-included theory for large $t-T_A$, 
which means that such theories with complex action and functional integral of future time 
are not excluded. 
Incidentally, as for the Hamiltonians in the future-included theory, 
there are two Hamiltonians $\hat{H}$ and $\hat{H}_B$, but  
only $\hat{H}$ is derived in ref.\cite{Nagao:2011is}. 
Therefore, in this paper we give the explicit derivation of $\hat{H}_B$ 
via the path integral using the method in ref.\cite{Nagao:2011is}, 
and confirm that it is given by $\hat{H}_B=\hat{H}^\dag$.


This paper is organized as follows. 
In section 2 we review our complex coordinate formalism and give a theorem 
for matrix elements. 
In section 3 we review the proper inner product for the Hamiltonian $\hat{H}$, 
and introduce another proper inner product for the Hamiltonian $\hat{H}_B$. 
Next we review the mechanism for suppressing the anti-Hermitian part of $\hat{H}$ 
after a long time development. 
In section 4 we study the various properties of the expectation value $\langle {\cal O} \rangle^{BA}$. 
We show that it allows us to have the Heisenberg equation, Ehrenfest's theorem and 
a conserved probability current density. 
In section 5 after reviewing the study in ref.\cite{Bled2006}, 
we show that the expectation value of 
the future-included theory for large $T_B -t$ and large $t-T_A$ 
corresponds to that of  
the future-not-included theory for large $t-T_A$ with a proper inner product. 
Section 6 is devoted to the summary and outlook. 
In appendix A we give an explicit derivation of $\hat{H}_B$ via the path integral 
following ref.\cite{Nagao:2011is}.

\section{Review of the complex coordinate formalism}\label{fundamental}


In this section we briefly review the complex coordinate formalism that we proposed 
in ref.\cite{Nagao:2011za} 
so that we can deal with the complex coordinate $q$ and momentum $p$ 
properly in the CAT. 
We emphasize that even in a real action theory (RAT) we encounter complex $q$ and $p$ 
at the saddle point in the cases of tunneling effect or WKB approximation, etc. 
As a simple and clear example, let us consider a wave function, 
\begin{equation} 
\psi(q) = \langle q | \psi \rangle. \label{psiq}
\end{equation}
This is defined for real $q$, but what happens for complex $q$ in the cases mentioned above? 
There are no problems with the left-hand side, because we can just say that the function $\psi$ is analytically 
extended to complex $q$. 
But the right-hand side cannot be extended to complex $q$, because $\langle q |$ 
is defined only for real $q$. 
Indeed $\langle q |$ obeys $\langle q | \hat{q} = \langle q | q$, 
so if we attempt to extend the real eigenvalue $q$ to complex, we encounter 
a contradiction because $\hat{q}$ is a Hermitian operator. 
Therefore $\hat{q}$ and $\langle q |$ 
have to be appropriately extended 
to a non-Hermitian operator and its eigenstate for complex $q$.

\subsection{Non-Hermitian operators $\hat{q}_{new}$ and $\hat{p}_{new}$, 
and the eigenstates of their Hermitian conjugates $|q \rangle_{new}$ and $|p \rangle_{new}$}

Following ref.~\cite{Nagao:2011za}, we summarize 
the construction of the non-Hermitian operators  of coordinate and momentum, 
$\hat{q}_{new}$ and $\hat{p}_{new}$, 
and the eigenstates  of their Hermitian conjugates 
$| q \rangle_{new}$ and $| p \rangle_{new}$ such that 
\begin{eqnarray}
&&\hat{q}_{new}^\dag  | q \rangle_{new} =q | q \rangle_{new} , \label{qhatqket=qqket_new} \\
&&\hat{p}_{new}^\dag  | p \rangle_{new} =p | p \rangle_{new} , \label{phatpket=ppket_new} \\
&&[\hat{q}_{new} , \hat{p}_{new}  ] = i \hbar , \label{comqhatphat}
\end{eqnarray}
for complex $q$ and $p$ by formally utilizing two coherent states. 
Our proposal is to replace 
$\hat{q}$, $\hat{p}$, $|q  \rangle$ and $|p  \rangle$ 
with $\hat{q}_{new}^\dag$, $\hat{p}_{new}^\dag$,  
$|q \rangle_{new}$ and $|p \rangle_{new}$. 
The explicit expressions for $\hat{q}_{new}$, $\hat{p}_{new}$, 
$| q \rangle_{new}$ and $| p \rangle_{new}$ are given by  
\begin{eqnarray}
&&\hat{q}_{new} \equiv 
\frac{1}{ \sqrt{1 - \frac{m' \omega'}{m\omega} }  } \left( \hat{q} - i \frac{\hat{p}}{m\omega} \right), \label{def_qhat_new} \\
&&\hat{p}_{new} \equiv 
\frac{1}{ \sqrt{1 - \frac{m' \omega'}{m\omega} }  }  \left( \hat{p} - \frac{m' \omega'}{i} \hat{q} \right) , \label{def_phat_new} \\
&&| q \rangle_{new} 
\equiv 
\left\{ \frac{m\omega}{4\pi \hbar} \left( 1 - \frac{m'\omega'}{m\omega} \right) \right\}^\frac{1}{4} 
e^{- \frac{m\omega}{4\hbar} \left( 1 - \frac{m'\omega'}{m\omega} \right) {q}^2 }
| \sqrt{ \frac{m\omega}{2\hbar} 
\left( 1 - \frac{m'\omega'}{m\omega} \right) } q \rangle_{coh} ,  \\
&&| p \rangle_{new} 
\equiv 
\left( \frac{1 - \frac{m'\omega'}{m\omega} }{4\pi \hbar m' \omega' } \right)^{\frac{1}{4}} 
e^{ -\frac{1}{4 \hbar m' \omega'}  \left( 1 - \frac{m'\omega'}{m\omega} \right)  p^2 }
| i \sqrt{ \frac{1}{2\hbar m' \omega'} 
\left( 1 - \frac{m'\omega'}{m\omega} \right)}  p \rangle_{coh'} , \label{defpketnew}
\end{eqnarray}
where $| \lambda \rangle_{coh}$ is a coherent state parametrized with a complex parameter $\lambda$ defined up to a normalization factor by 
$| \lambda \rangle_{coh} 
\equiv  
e^{\lambda a^\dag} | 0 \rangle 
= \sum_{n=0}^{\infty} \frac{\lambda^n}{\sqrt{n!}} | n \rangle$, 
and this satisfies the relation 
$a | \lambda \rangle_{coh} = \lambda | \lambda \rangle_{coh}$. 
Here 
$a = \sqrt{ \frac{m\omega}{2\hbar}}  \left( \hat{q} + i \frac{ \hat{p}}{m \omega}  \right)$ and 
$a^\dag = \sqrt{ \frac{m\omega}{2\hbar}}\left( \hat{q} - i \frac{ \hat{p}}{m \omega}  \right)$
are annihilation and creation operators, 
where $\hat{q}$ and $\hat{p}$ are the usual Hermitian  
operators of coordinate and momentum obeying 
\begin{eqnarray}
&&\hat{q} | q \rangle = q| q \rangle , \label{qhatqket=qqket} \\
&&\hat{p} | p \rangle = p | p \rangle ,    \label{phatpket=ppket} \\ 
&&[ \hat{q},  \hat{p} ] = i \hbar \label{commu_rel_qhatphat}
\end{eqnarray}
for real $q$ and $p$. 
In eq.(\ref{defpketnew}) 
$| \lambda \rangle_{coh'}$ is another coherent state, which is defined 
similarly with different parameters $m'\omega'$, 
$| \lambda \rangle_{coh'} \equiv e^{\lambda {a'}^\dag} | 0 \rangle$, 
where ${a'}^\dag$  is given by 
${a'}^\dag = \sqrt{ \frac{m'\omega'}{2\hbar}}
\left( \hat{q} - i \frac{ \hat{p}}{m' \omega'}  \right) \label{creation'}$. 
Before seeing the properties of $\hat{q}_{new}$, $\hat{p}_{new}$, 
$| q \rangle_{new}$, and $| p \rangle_{new}$, 
we define a delta function of complex parameters in the next subsection.

\subsection{The delta function}\label{subsec:deltafunc}

\begin{figure}[htb]
\begin{center}
\includegraphics[height=10cm]{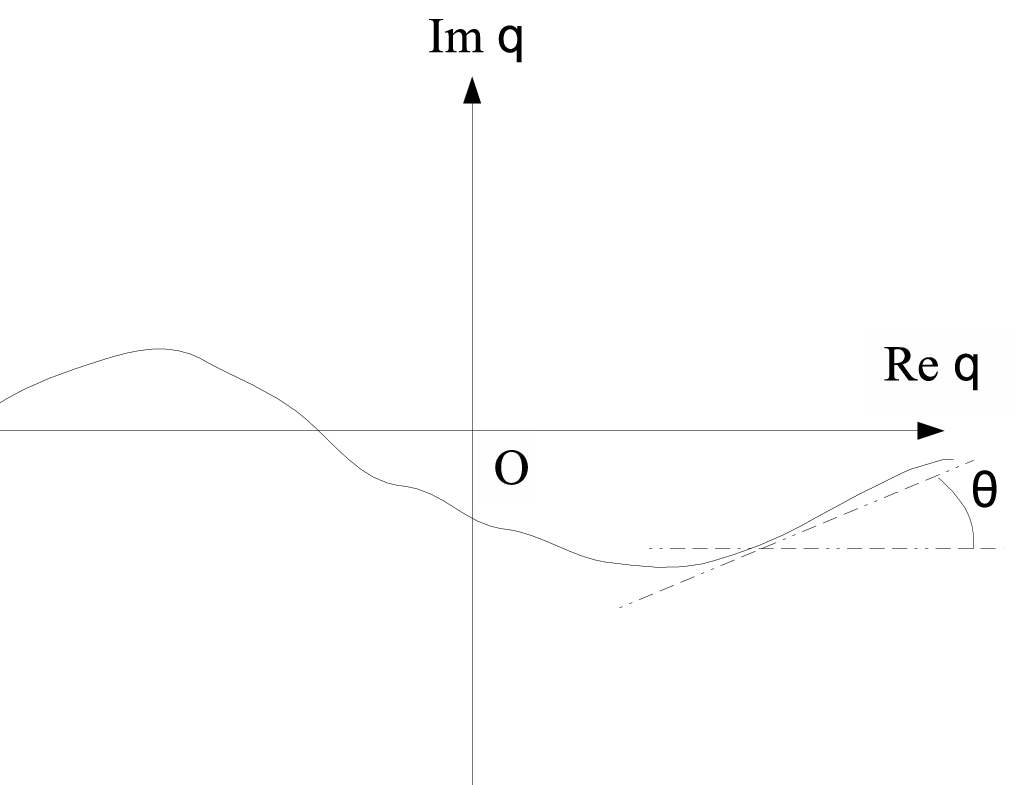}
\end{center}
\caption{An example of permitted paths}
\label{fig:contour}
\end{figure}

\begin{figure}[htb]
\begin{center}
\includegraphics[height=10cm]{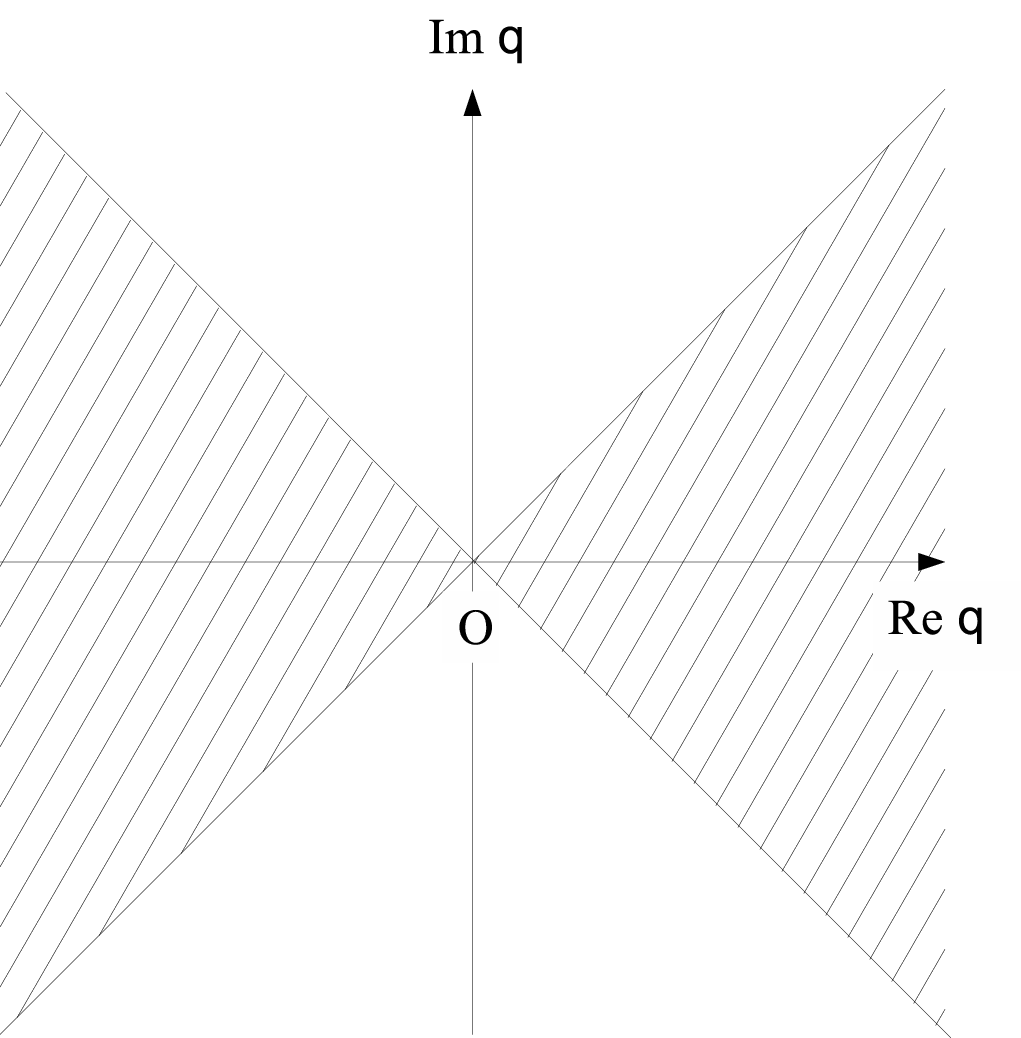}
\end{center}
\caption{Domain of the delta function}
\label{fig:delta_function}
\end{figure}

For our later convenience we first define ${\cal D}$ as 
a class of distributions depending on one complex variable $q \in \mathbf{C}$. 
Using a function $g:{\mathbf C} \rightarrow {\mathbf C}$ as 
a distribution in the class ${\cal D}$, 
we define the following functional $G$ 
\begin{equation}
G[f] = \int_C f(q) g(q) dq  \label{mappingG}
\end{equation}
for any analytical function $f:{\mathbf C} \rightarrow {\mathbf C}$ 
with convergence requirements such that $f \rightarrow 0$ for $q \rightarrow \pm \infty$. 
The functional $G$ is a linear mapping from the function $f$ to a complex number. 
Since the simulated function $g$ is supposed to be analytical in $q$, 
the path $C$, which is chosen 
to run from $-\infty$ to $\infty$ in the complex plane, 
can be deformed freely and so it is not relevant. 
As an approximation to such a distribution we could use 
the smeared delta function defined for complex $q$ by 
\begin{equation}
g(q) = \delta_c^\epsilon(q) 
\equiv \sqrt{\frac{1}{4 \pi \epsilon}} e^{-\frac{q^2}{4\epsilon}} , \label{delta_c_epsilon(q)}
\end{equation}
where $\epsilon$ is a finite small positive real number. 
For the limit of $\epsilon \rightarrow 0$ $g(q)$ converges in the distribution sense 
for complex $q$ obeying the condition 
\begin{equation}
L(q) 
\equiv \left( \text{Re}(q) \right)^2 - \left( \text{Im}(q) \right)^2 
>0 . \label{cond_of_q_for_delta} 
\end{equation}
For any analytical test function $f(q)$\footnote{Because of the Liouville theorem if $f$ is 
a bounded entire function, 
$f$ is constant. So we are considering as $f$ an unbounded entire function or a function 
that is not entire but is holomorphic at least in the region on which the path runs.} 
and any complex $q_0$ this $\delta_c^\epsilon(q)$ satisfies 
\begin{equation}
\int_C f(q) \delta_c^\epsilon(q-q_0) dq = f(q_0) ,
\end{equation}
as long as we choose the path $C$ 
such that it runs from $-\infty$ to $\infty$ in the complex plane 
and at any $q$ its tangent line and a horizontal line 
form an angle $\theta$ whose absolute value  
is within $\frac{\pi}{4}$ to satisfy the inequality (\ref{cond_of_q_for_delta}). 
An example of permitted paths is shown in fig.\ref{fig:contour}, 
and the domain of the delta function is drawn in fig.\ref{fig:delta_function}. 
%


Next we extend the delta function to complex $\epsilon$, 
and consider 
\begin{eqnarray}
\delta_c^{\epsilon}(aq) 
&=& \sqrt{\frac{1}{4 \pi \epsilon}} e^{-\frac{1}{4\epsilon} a^2 q^2} 
\label{deltaepsilonaq}
\end{eqnarray}
for non-zero complex $a$. 
We express $\epsilon$, $q$, and $a$ as 
$\epsilon = r_\epsilon e^{i\theta_\epsilon}$, 
$q= r e^{i\theta}$, and $a = r_a e^{i\theta_a}$. 
The convergence condition of $\delta_c^{\epsilon}(aq)$: 
$\text{Re} \left( \frac{a^2 q^2}{\epsilon} \right) > 0$ 
is expressed as 
\begin{eqnarray}
&&-\frac{\pi}{4} + \frac{1}{2} ( \theta_{\epsilon} - 2\theta_a ) 
< \theta < \frac{\pi}{4} + \frac{1}{2} ( \theta_{\epsilon}  - 2 \theta_a ), 
\label{cond1forqa} \\
&&\frac{3}{4} \pi + \frac{1}{2}( \theta_{\epsilon} - 2 \theta_a ) 
< \theta < \frac{5}{4}\pi + \frac{1}{2} ( \theta_{\epsilon} - 2 \theta_a ) . 
\label{cond2forqa}
\end{eqnarray}
For $q$, $\epsilon$, and $a$ such that eqs.(\ref{cond1forqa})(\ref{cond2forqa}) 
are satisfied, $\delta_c^{\epsilon}(aq)$ behaves well as a delta function 
of $aq$, and we obtain the relation 
\begin{equation}
\delta_c^{\epsilon}(aq) = 
\frac{\text{sign}(\text{Re} a)}{ a } \delta_c^{\frac{\epsilon}{a^2}}(q) , 
\label{scaling_deltafunction} 
\end{equation}
where we have introduced an expression 
\begin{eqnarray}
\text{sign}(\text{Re} a) 
\equiv 
\left\{ 
 \begin{array}{cc}
1  & \text{for} ~\text{Re}(a) > 0 , \\
-1 & \text{for} ~\text{Re}(a) < 0  . \\ 
 \end{array}
\right.
\end{eqnarray}
%

\subsection{New devices to handle complex parameters}\label{newdevices}

In this subsection, to keep the analyticity in dynamical variables of FPI such as 
$q$ and $p$ 
we define a modified set of a complex conjugate, 
Hermitian conjugates, and bras.

\subsubsection{Modified complex conjugate $*_{ \{ \} }$}

We define a modified complex conjugate for a function of $n$-parameters 
$f( \{a_i \}_{i=1, \ldots, n} )$ by 
\begin{equation}
f(\{a_i \}_{i=1, \ldots, n} )^{*_{\{a_i | i \in A \}} } = f^*( \{a_i \}_{i \in A}  ,  \{a_i ^*\}_{i \not\in A} ) , 
\end{equation}
where $A$ denotes the set of indices attached to the parameters 
in which we keep the analyticity, 
and, on the right-hand side, $*$ on $f$ acts on the coefficients included in $f$. 
For example, the complex conjugates $*_q$ and $*_{q,p}$ 
of a function $f(q,p)=a q^2 + b p^2$ are  
$f(q,p)^{*_q} = a^* q^2 + b^* (p^*)^2$ and 
$f(q,p)^{*_{q,p}} = a^* q^2 + b^* p^2$.
The analyticity is kept 
in $q$, and both $q$ and $p$, respectively. 
For simplicity we express the modified complex conjugate as $*_{ \{  \} }$.

\subsubsection{Modified bras ${}_m \langle ~|$ and ${}_{ \{ \} } \langle ~|$, 
and modified Hermitian conjugate $\dag_{ \{ \} }$ }

For some state $| \lambda \rangle$ with some complex parameter $\lambda$, 
we define a modified bra ${}_m\langle \lambda |$ by 
\begin{equation} 
{}_m\langle \lambda | \equiv \langle \lambda^* |  \label{modified_bra_anti-linear}
\end{equation}
so that it preserves the analyticity in $\lambda$. 
In the special case of $\lambda$ being real it becomes a usual bra. 
In addition we define a slightly generalized modified bra 
${}_{\{\}}\langle ~|$ and a modified Hermitian conjugate $\dag_{ \{ \} }$ of a ket, 
where $\{ \}$ is a symbolical expression for a set of parameters 
in which we keep the analyticity. 
For example, ${}_{u,v}\langle u | = {}_u \langle u | = {}_m\langle u |$, 
$( | u \rangle )^{\dag_{u, v}}  =( | u \rangle )^{\dag_{u}}  = {}_m \langle u |$. 
We express the Hermitian conjugate $\dag_{ \{ \} }$ of a ket symbolically as 
$( |  ~\rangle )^{\dag_{\{ \} }} = {}_{\{ \}}\langle  ~|$. 
Also, we write the Hermitian conjugate $\dag_{ \{ \} }$ of a bra as 
$( {}_{\{ \}}\langle  ~| )^{\dag_{\{ \} }} =  |  ~\rangle$. 
So for a matrix element we have the relation 
${}_{\{\}}\langle u | A | v \rangle^{*_{ \{ \} }} = 
{}_{ \{ \} } \langle v | A^\dag | u \rangle$.

\subsection{Properties of $\hat{q}_{new}$, $\hat{p}_{new}$, $|q \rangle_{new}$ and $|p \rangle_{new}$, 
and a theorem for matrix elements}

The states $| q \rangle_{new}$ and $| p \rangle_{new}$ are normalized 
so that they satisfy 
the following relations, 
\begin{eqnarray}
{}_m\langle_{new}~ q' | q \rangle_{new} 
&=& \delta_c^{\epsilon_1} ( q' - q ) , \label{m_q'branew_qketnew}\\ 
{}_m\langle_{new}~ p' | p \rangle_{new} 
&=& \delta_c^{\epsilon'_1} ( p' - p ) , \label{m_p'branew_pketnew}
\end{eqnarray}
where 
$\epsilon_1 = \frac{\hbar}{m \omega \left( 1 - \frac{m'\omega'}{m\omega} \right)}$ 
and 
$\epsilon'_1 = \frac{\hbar m'\omega'}{1 - \frac{m'\omega'}{m\omega} }$. 
For sufficiently large $m\omega$ and small $m'\omega'$ 
the delta functions converge 
for complex $q$, $q'$, $p$, and $p'$ satisfying the conditions 
$L(q-q') > 0$ and $L(p-p') > 0$, where $L$ is given in eq.(\ref{cond_of_q_for_delta}). 
These conditions are satisfied only when $q$ and $q'$ or $p$ and $p'$ are on the same paths 
respectively. 
In the following we take $m\omega$ sufficiently large and $m'\omega'$ sufficiently small. 
Then eqs.(\ref{m_q'branew_qketnew})(\ref{m_p'branew_pketnew}) represent 
the orthogonality relations for $| q \rangle_{new}$ and $| p \rangle_{new}$, 
and we have the following relations for complex $q$ and $p$: 
\begin{eqnarray}
&&\int_C dq | q \rangle_{new} ~{}_m \langle_{new} q |  = 1 , \label{completion_complexq_ket2} \\
&&\int_C dp | p \rangle_{new} ~{}_m \langle_{new} p |  = 1 , \label{completion_complexp_ket2} \\
&&\hat{p}_{new}^\dag | q \rangle_{new} 
=i \hbar \frac{\partial}{\partial q} | q \rangle_{new}, \label{phatnewqketnew2} \\
&&\hat{q}_{new}^\dag | p \rangle_{new} 
= \frac{\hbar}{i} \frac{\partial}{\partial p} | p \rangle_{new}, \label{qhatnewpketnew2} \\
&&{}_m\langle_{new}~ q | p \rangle_{new} 
= \frac{1}{\sqrt{2 \pi \hbar}} \exp\left(\frac{i}{\hbar}p q \right). 
\label{basis_fourier_transf2}  
\end{eqnarray}
Thus $\hat{q}_{new}^\dag$, $\hat{p}_{new}^\dag$, 
$| q \rangle_{new}$ and $| p \rangle_{new}$ 
with complex $q$ and $p$ 
obey the same relations that $\hat{q}$, $\hat{p}$, $| q \rangle$, and $| p \rangle$ with 
real $q$ and $p$ satisfy. 
In the limits of $m\omega \rightarrow \infty$ and $m'\omega' \rightarrow 0$ 
$\delta_c^{\epsilon_1} ( q' - q )$, $\delta_c^{\epsilon'_1} ( p' - p )$ and 
$\exp\left( \frac{i}{\hbar}p q \right)$ 
in eqs.(\ref{m_q'branew_qketnew})(\ref{m_p'branew_pketnew})(\ref{basis_fourier_transf2}) 
are well-defined as distributions of the type ${\cal D}$, which is 
introduced in subsection~\ref{subsec:deltafunc}.

For real $q'$ and $p'$, $| q' \rangle_{new}$ and $| p' \rangle_{new}$ become 
$| q' \rangle$ and $| p' \rangle$ respectively. 
Also, for them, 
$\hat{q}_{new}^\dag$ and $\hat{p}_{new}^\dag$ behave like  
$\hat{q}$ and $\hat{p}$ respectively. 
In relation to the disappearance of the anti-Hermitian terms in 
$\hat{q}_{new}^\dag$ and $\hat{p}_{new}^\dag$, 
we put forward a theorem for matrix elements 
of the form 
${}_m\langle_{new}~ q' ~\text{or}~ p'|
{\cal O}( \hat{q}_{new}, \hat{q}_{new}^\dag, \hat{p}_{new}, \hat{p}_{new}^\dag) | q'' ~\text{or}~ p'' \rangle_{new}$, 
where 
${\cal O}( \hat{q}_{new}, \hat{q}_{new}^\dag, \hat{p}_{new}, \hat{p}_{new}^\dag)$ 
is a Taylor-expandable function of the four operators 
$\hat{q}_{new}$, $\hat{q}_{new}^\dag$, $\hat{p}_{new}$, and $\hat{p}_{new}^\dag$. 
We easily see that such a matrix element 
can be expressed as the summation of the products of factors made 
of $q'$, $p'$, $q''$, $p''$ 
or their differential operators 
and the distributions $\delta_c^{\epsilon_1}(q' - q'')$, 
$\delta_c^{\epsilon'_1}(p' - p'')$, or 
$\exp\left(\pm \frac{i}{\hbar} p' q'' \right)$. 
Then, since we shall extract only analytically 
weighted results from the matrix element, 
we do not have to worry about 
the anti-Hermitian terms in $\hat{q}_{new}$, $\hat{q}_{new}^\dag$, $\hat{p}_{new}$ 
and $\hat{p}_{new}^\dag$, 
provided that we are satisfied with the result in the distribution sense. 
So we pose the following theorem.

\vspace{0.5cm}
\noindent
Theorem: 
The matrix element 
${}_m\langle_{new}~ q' ~\text{or}~ p'|
{\cal O}( \hat{q}_{new}, \hat{q}_{new}^\dag, \hat{p}_{new}, \hat{p}_{new}^\dag) | q'' ~\text{or}~ p'' \rangle_{new}$   
can be evaluated as if inside the operator ${\cal O}$ 
we had the hermiticity conditions 
$\hat{q}_{new} \simeq \hat{q}_{new}^\dag \simeq \hat{q}$ and 
$\hat{p}_{new} \simeq  \hat{p}_{new}^\dag \simeq \hat{p}$ 
for $q'$, $q''$, $p'$, $p''$ such that the resulting quantities  
are well-defined in the sense of distribution.

\vspace{0.5cm}

\noindent
This theorem could help us from running into calculations that are too hard 
for the use of our complex coordinate formalism.

\subsection{Remarks on the complex coordinate formalism}

We have seen that $\hat{q}_{new}$ and $\hat{p}_{new}$ etc. have nice properties, but 
one might still feel a bit uneasy about our replacement of 
$\hat{q}$ and $\hat{p}$ with $\hat{q}_{new}$ and $\hat{p}_{new}$. 
To accept $\hat{q}_{new}$ and $\hat{p}_{new}$, 
it might help slightly to have in mind that 
operators smooth in $\hat{q}$ and $\hat{p}$ 
like $\hat{q}_{new}$ and $\hat{p}_{new}$ generically have eigenvalues 
filling the whole complex plane, 
while Hermitian operators like $\hat{q}$ and $\hat{p}$ 
have eigenvalues only along a certain curve, e.g., on the real axis in the complex plane. 
For our purpose of having general contours running through eigenvalues 
we replaced the special operators $\hat{q}$ and $\hat{p}$ 
by the more generic ones $\hat{q}_{new}$ and $\hat{p}_{new}$. 
The philosophy should be that almost any small disturbance would anyway 
bring $\hat{q}$ and $\hat{p}$ 
into operators of the generic type with the whole complex plane as a spectrum. 
The operators $\hat{q}_{new}$ and $\hat{p}_{new}$ are just concrete examples of 
such tiny deformation.
So we stress that the Hermitian operators as $\hat{q}$ and $\hat{p}$ are special by 
having their eigenvalue spectrum on a curve, e.g., on the real axis in the complex plane 
rather than distributed all over it. 
If we had clung to the belief in curve-spectra, it would have been embarrassing 
for our formalism that 
under Heisenberg time development one could have feared that, 
from time to time in our scheme, 
the curve-spectra 
would be transformed into new curve-spectra that might not match 
at the free contour choice. 
Now, however, 
as already stressed, if we use $\hat{q}_{new}$ and $\hat{p}_{new}$, 
from the beginning 
we have already gone over to operators with any complex numbers as eigenvalues. 
So arbitrary deformation of the contour would a priori cause no problems. 
Thus we claim that the contours of integration can be chosen 
freely at each time $t$, so that there is no need for any natural choice, 
which only has to run from $-\infty$ to $\infty$. 

We come back to the problem we raised at the beginning of this section: 
how eq.(\ref{psiq}) is expressed for complex $q$. 
Now we can express it based on our complex coordinate formalism as 
\begin{equation}
\psi(q) = {}_m\langle_{new}~ q | \psi \rangle . \label{psiqnew}
\end{equation}
When $q$ is real, this becomes the usual expression of eq.(\ref{psiq}). 
In addition we mention another problem on the momentum relation $p= m \dot{q}$ 
in the CAT, where the mass $m$ is generically complex. 
Indeed we encounter a contradiction if we cling to 
the real eigenvalues $q$ and $p$ of 
the usual Hermitian operators $\hat{q}$ and $\hat{p}$.  
In ref.\cite{Nagao:2011is}, 
we have explicitly examined the definitions of the momentum and Hamiltonian via FPI, 
and confirmed that they have the same forms as those in the RAT but 
with complex eigenvalues $q$ and $p$ of non-Hermitian operators $\hat{q}_{new}$ 
and $\hat{p}_{new}$. 
In this section, we have briefly reviewed the complex coordinate formalism; 
for details see ref.\cite{Nagao:2011za}. 
Finally, we show a summary of the comparison between the RAT and the CAT 
in Table \ref{tab1:comparison}. 
\begin{table}
\caption[AAA]{Various quantities in the RAT and the CAT}
\label{tab1:comparison}
\begin{center}
\begin{tabular}{|p{4cm}|p{4cm}|p{6.5cm}|}
\hline
    & the RAT & the CAT \\
\hline
parameters  &    $q$, $p$ real,   & $q$, $p$ complex \\  
\hline
complex conjugate  &    $*$   &  $*_{ \{~\}}$  \\  
\hline
Hermitian conjugate  &    $\dag$   &  $\dag_{ \{~\}}$  \\  
\hline
delta function of $q$ &  $\delta(q)$ defined for    & $\delta_c(q)$ defined for $q$ s.t.\\
& real $q$ &  $\left( \text{Re}(q) \right)^2 > \left( \text{Im}(q) \right)^2$  \\ 
\hline
bras of $| q\rangle$, $| p\rangle$   &  $\langle q | = ( | q\rangle )^\dag $,  & 
${}_m\langle_{new}~ q |=\langle_{new}~ q^* |  = ( | q \rangle_{new} )^{\dag_q}  $,  \\
& $\langle p | = ( | p\rangle )^\dag$ & ${}_m\langle_{new}~ p |=\langle_{new}~ p^* |  = ( | p \rangle_{new} )^{\dag_p}  $ \\

\hline  
completeness for  &  $\int_{-\infty}^\infty |q\rangle \langle q | dq =1$ , & 
$\int_C |q \rangle_{new} ~{}_m\langle_{new}~ q | dq =1$ , \\
$| q \rangle$ and $|p\rangle$ &  $\int_{-\infty}^\infty |p \rangle \langle p | dp =1$  & 
$\int_C |p\rangle_{new} ~{}_m\langle_{new}~ p | dp =1$ \\
& along real axis & $C$: any path running from $-\infty$ to $\infty$\\
\hline 
orthogonality for  &  $\langle q | q'\rangle = \delta(q-q')$ , & 
${}_m\langle_{new}~ q | q'\rangle_{new} = \delta_c^{\epsilon_1} (q-q')$ , \\
$| q\rangle$ and $|p\rangle$ &  $\langle p | p'\rangle = \delta(p-p')$ & 
${}_m\langle_{new}~ p | p'\rangle_{new} = \delta_c^{\epsilon'_1} (p-p')$ \\
\hline 
basis of Fourier expansion &  $\langle q | p\rangle = \exp(ipq)$ & 
${}_m\langle_{new}~ q | p\rangle_{new} = \exp(ipq)$ \\
\hline 
$q$ representation of  $| \psi \rangle$  & $\psi(q) = \langle q | \psi \rangle$ & 
$\psi(q) = {}_m\langle_{new}~ q | \psi \rangle $ \\
\hline 
complex conjugate of  $\psi(q)$ & $\langle q | \psi \rangle^*=\langle \psi | q \rangle$ & 
${}_m\langle_{new}~ q | \psi \rangle^{*_q } =\langle \psi | q \rangle_{new}$ \\
\hline
normalization of $\psi(q)$  & $\int_{-\infty}^\infty \psi(q)^* \psi(q) dq = 1$ & 
$\int_C \psi(q)^{*_q } \psi(q) dq = 1$ \\
\hline
\end{tabular}
\end{center}
\end{table}
%
%

\section{Proper inner products and the automatic hermiticity mechanism}

We begin with an explanation of the two states $\langle B(t) |$ and $| A(t) \rangle$, 
definable from their respective functional integrals of future and past 
following ref.\cite{Bled2006}, and 
review the proper inner product for the Hamiltonian $\hat{H}$ 
in the future-not-inclulded theory 
following ref.\cite{Nagao:2010xu,Nagao:2011za}. 
Then we construct the proper inner product for the other Hamiltonian $\hat{H}_B$ 
in the future-included theory. 
Furthermore we review the mechanism\cite{Nagao:2010xu,Nagao:2011za} 
for suppressing the anti-Hermitian part of the Hamiltonian.

\subsection{Definitions of $| A(t) \rangle$ and $\langle B(t) |$}

In ref.\cite{Bled2006}, the state $| A(t) \rangle$ is introduced by 
$\langle q | A(t) \rangle = \int_{\text{path}(t)=q} 
e^{\frac{i}{\hbar} S_{-\infty ~\text{to} ~t}} D\text{path}$ , 
where $\text{path}(t)=q$ means the boundary condition at the time $t$. 
We write this in a slightly modified form as 
\begin{equation}
\psi_A(q)={}_m \langle_{new}~ q | A(t) \rangle = \int_{\text{path}(t)=q} 
e^{\frac{i}{\hbar} S_{T_A ~\text{to} ~t}} D\text{path}   \label{psi_A(q)}
\end{equation}
based on the complex coordinate formalism\cite{Nagao:2011za} 
so that it is properly defined even for complex $q$. 
The other state $| B(t) \rangle$ is introduced in ref.\cite{Bled2006} as 
$\langle B(t) | q \rangle \equiv \int_{\text{path}(t)=q} e^{\frac{i}{\hbar} S_{t ~\text{to} ~\infty}} D\text{path}$, 
but we express it as 
\begin{equation}
\psi_B(q)^{*_q} = \langle B(t) | q \rangle_{new} 
= \int_{\text{path}(t)=q} e^{\frac{i}{\hbar} S_{t ~\text{to} ~\infty}} D\text{path} ,  
\end{equation}
by defining $\psi_B(q)$ by 
\begin{equation}
\psi_B(q) = ~{}_m \langle_{new}~ q | B \rangle .  \label{psi_B(q)} 
\end{equation}
Here 
$\psi_A(q)$ and $\psi_B(q)$ are kinds of wave functions of the universe at time $t$, 
which include information on the past and future times, respectively. 
The states $| A(t) \rangle$ and $| B(t) \rangle$ time-develop according to 
\begin{eqnarray}
&&i \hbar \frac{d}{dt} | A(t) \rangle = \hat{H} | A(t) \rangle , \label{schro_eq_Astate} \\
&&i \hbar \frac{d}{dt} | B(t) \rangle = \hat{H}_B | B(t) \rangle , \label{schro_eq_Bstate} 
\end{eqnarray}
where 
\begin{equation}
\hat{H}_B= \hat{H}^\dag . \label{defHB}
\end{equation}
We note that in ref.\cite{Nagao:2011is} we explicitly derived the form of $\hat{H}$ 
-- in a system with a single degree of freedom for simplicity -- 
via the Feynman path integral as follows: 
\begin{equation}
\hat{H}= \frac{1}{2m} (\hat{p}_{new})^2 + V(\hat{q}_{new}) , \label{expHhat}
\end{equation}
where we supposed that 
$V(\hat{q}_{new}) = \sum_{n = 2}^\infty b_n (\hat{q}_{new})^n$. 
This is the first application of the complex coordinate formalism.  
In appendix~\ref{der_H_B} we derive $\hat{H}_B$ in a similar way.

\subsection{A proper inner product for $\hat{H}$}

As preparation for deriving the proper inner product for $\hat{H}_B$ 
we briefly review and explain the construction of the proper inner product for $\hat{H}$, 
which we introduced in refs.\cite{Nagao:2010xu, Nagao:2011za} 
in a future-not-included theory. 
To explain it we introduce the eigenstates $| \lambda_i \rangle (i=1,2,\cdots)$ of 
the Hamiltonian $\hat{H}$ obeying 
$\hat{H} | \lambda_i \rangle = \lambda_i | \lambda_i \rangle$, 
where $\lambda_i (i=1,2,\cdots)$ are the eigenvalues of $\hat{H}$, 
and define the diagonalizing operator $P$ of the Hamiltonian $\hat{H}$ by 
$P=(| \lambda_1 \rangle , | \lambda_2 \rangle , \ldots)$. 
Then $\hat{H}$ is diagonalized as 
$\hat{H} = PD P^{-1}$, 
where $D$ is given by $\text{diag}(\lambda_1, \lambda_2, \cdots)$. 
We introduce an orthonormal basis $| e_i \rangle (i=1, \ldots)$ satisfying 
$\langle e_i | e_j \rangle = \delta_{ij}$ by 
$D | e_i \rangle = \lambda_i   | e_i \rangle$. 
The basis $| e_i \rangle$ is related to $| \lambda_i \rangle$ as 
$| \lambda_i \rangle = P | e_i \rangle$. 
We note that the $| \lambda_i \rangle$ are not orthogonal to each other 
in the usual inner product $I$, 
$I(| \lambda_i \rangle , | \lambda_j \rangle ) = \langle \lambda_i | \lambda_j \rangle \neq \delta_{ij}$. 
So the theory defined with $I$ would measure unphysical transitions. 
To make a physically reasonable measurement, 
in refs\cite{Nagao:2010xu, Nagao:2011za} 
we introduced a proper inner product $I_Q$ for arbitrary kets $|u \rangle$ 
and $|v \rangle$ as
\begin{equation}
I_Q(|u \rangle , |v \rangle) = \langle u |_Q v \rangle 
= \langle u | Q | v \rangle ,  
\end{equation}
where $Q$ is a Hermitian operator chosen as 
\begin{equation}
Q=(P^\dag)^{-1} P^{-1}  \label{def_QA}
\end{equation}
so that the eigenstates of $\hat{H}$ become orthogonal to each other with regard to $I_Q$, 
$I_Q( | \lambda_i \rangle , |\lambda_j \rangle) = \delta_{ij}$. 
This enables us to make a physically reasonable observation, and also implies 
the orthogonality relation 
$\sum_i | \lambda_i \rangle \langle \lambda_i |_{Q} = 1$. 
We note that $I_Q$ is different from the CPT inner product defined 
in the PT symmetric Hamiltonian formalism\cite{PTsym_Hamiltonians}.

Via the inner product $I_Q$ we define the $Q$-Hermitian conjugate of some operator $A$ by 
$\langle \psi_2 |_Q A | \psi_1 \rangle^* = \langle \psi_1 |_Q A^{\dag^Q} | \psi_2 \rangle$, 
from which we see that $A^{\dag^Q}$ is written as 
$A^{\dag^Q} = Q^{-1} A^\dag Q$. 
Similarly, in the case where $|\psi_1 \rangle$ or $|\psi_2 \rangle$ are given in states 
as $|u \rangle$ or $|v \rangle$, 
we can consider the following relation: 
${}_{\{ \}}\langle v |_Q A | u \rangle^{*_{\{ \}}} = 
{}_{\{ \}}\langle u |_Q A^{\dag^Q} | v \rangle$, 
where $\{ ~\}$ denotes a set of parameters in which we keep the analyticity. 
We also define $\dag^Q$ for kets and bras as 
$| \lambda \rangle^{\dag^Q} \equiv \langle \lambda |_Q $ and 
$\left(\langle \lambda |_Q \right)^{\dag^Q} \equiv | \lambda \rangle$.  
Similarly, we define $\dag^Q_{\{ \}}$ for kets and bras as 
$| \lambda \rangle^{\dag^Q_{ \{ \} }} \equiv {}_{ \{ \} }\langle \lambda |_Q $ and 
$\left( {}_{ \{ \} } \langle \lambda |_Q \right)^{\dag^Q_{ \{ \} }} \equiv | \lambda \rangle$.  
When some operator $A$ satisfies 
$A^{\dag^Q} = A$ , 
we call $A$ $Q$-Hermitian. 
This is the definition of $Q$-hermiticity. 
\footnote{We note that in ref.\cite{Geyer} a similar inner product 
was studied and a criterion for identifying a unique
inner product through the choice of physical observables was also provided.}

Furthermore we explain the $Q$-normality of $H$.   
Since 
\begin{equation}
``P^{\dag^Q}"
\equiv
\left(
 \begin{array}{c}
      \langle \lambda_1 |_Q     \\
      \langle \lambda_2 |_Q     \\
      \vdots 
 \end{array}
\right) = P^{-1} 
\end{equation}
satisfies 
$``P^{\dag^Q}" \hat{H} P = D$ and $``P^{\dag^Q}" \hat{H}^{\dag^Q} P = D^{\dag}$, 
$\hat{H}$ is $Q$-normal, 
$[\hat{H}, \hat{H}^{\dag^Q} ] = P [D, D^\dag ] P^{-1} =0$.  
In other words the inner product $I_Q$ is defined so that 
$\hat{H}$ is normal with regard to it.

\subsection{A proper inner product for $\hat{H}_B$}

Following the construction of the proper inner product for $\hat{H}$ 
in the previous subsection 
we construct the proper inner product $I_{Q_B}$ for $\hat{H}_B$. 
Taking the Hermitian conjugate of the relation 
$\langle \lambda_j |_Q \hat{H} = \lambda_j \langle \lambda_j |_Q$ , 
we obtain 
$\hat{H}^\dag Q | \lambda_j \rangle = \lambda_j^* Q | \lambda_j \rangle$. 
Using eq.(\ref{defHB}) we rewrite this as 
\begin{equation}
\hat{H}_B | \lambda_j \rangle_B = \lambda_j^* | \lambda_j \rangle_B , \label{HBlambdatildeket}
\end{equation}
where we have introduced 
$| \lambda_j \rangle_B \equiv Q| \lambda_j \rangle$. 
Thus the eigenstates and the eigenvalues of $\hat{H}_B$ are given by 
$| \lambda_j \rangle_B$ and $\lambda_j^* ~(j=1,2,\ldots)$ respectively, 
and the diagonalizing matrix of $\hat{H}_B$ is given by 
$P_B \equiv (| \lambda_1 \rangle_B , | \lambda_2 \rangle_B , \ldots) 
=Q P = (P^\dag)^{-1}$. 
We introduce a proper inner product $I_{Q_B}$ 
for arbitrary kets $|u \rangle$ and $|v \rangle$ as 
$I_{Q_B}(|u \rangle , |v \rangle) = \langle u |_{Q_B} v \rangle 
= \langle u | {Q_B} | v \rangle$, 
where $Q_B$ is a Hermitian operator chosen as 
\begin{equation}
Q_B
= (P_B^\dag)^{-1} P_B^{-1} 
= Q^{-1}   \label{QB_rel_Q-1}
\end{equation}
in order that $| \lambda_j \rangle_B$ become orthogonal to each other with regard to $I_{Q_B}$, 
$I_{Q_B}( | \lambda_i \rangle_B , | \lambda_j \rangle_B) = \delta_{ij}$. 
Then we also have the completeness relation 
$\sum_i | \lambda_i \rangle_B ~{}_B\langle \lambda_i |_{Q_B} = 1$.

Taking the $Q_B$-Hermitian conjugate of eq.(\ref{HBlambdatildeket}), we obtain 
${}_B\langle \lambda_i |_{Q_B} \hat{H}_B^{\dag^{Q_B}} = {}_B\langle \lambda_i |_{Q_B} \lambda_i$, 
where $\hat{H}_B^{\dag^{Q_B}}$ is given by 
\begin{equation}
\hat{H}_B^{\dag^{Q_B}} = Q_B^{-1}  \hat{H}_B^\dag Q_B . \label{HBdagQB}
\end{equation}
Since 
\begin{equation}
``P_B^{\dag^{Q_B}}"
\equiv
\left(
 \begin{array}{c}
      {}_B \langle \lambda_1 |_{Q_B}     \\
      {}_B \langle \lambda_2 |_{Q_B}     \\
      \vdots 
 \end{array}
\right) = ( P_B )^{-1}   
\end{equation}
satisfies 
$``P_B^{\dag^{Q_B}}" \hat{H}_B P_B = D^\dag$ and  
$``P_B^{\dag^{Q_B}}" \hat{H}^{\dag^{Q_B}} P_B = D$, 
$\hat{H}_B$ is $Q_B$-normal, $[\hat{H}_B, \hat{H}_B^{\dag^{Q_B} }] =0$.

For later convenience we decompose $\hat{H}$ as 
$\hat{H}=\hat{H}_{Qh} + \hat{H}_{Qa}$, 
where $\hat{H}_{Qh}= \frac{\hat{H} + \hat{H}^{\dag^Q} }{2}$ and 
$\hat{H}_{Qa} = \frac{\hat{H} - \hat{H}^{\dag^Q} }{2}$ are 
$Q$-Hermitian and anti-$Q$-Hermitian parts of $\hat{H}$ respectively. 
We also decompose $D$ as $D=D_R + iD_I$, 
where we have introduced $D_R= \frac{D + D^\dag }{2}$ and $D_I= \frac{D - D^\dag }{2}$. 
The diagonal components of $D_R$ and $D_I$ are the real 
and imaginary parts of 
the diagonal components of $D$ respectively. 
Then $\hat{H}_{Qh}$ and $\hat{H}_{Qa}$ can be expressed in terms of $D_R$ and $D_I$ as 
$\hat{H}_{Qh} = P D_R P^{-1}$ and $\hat{H}_{Qa} = i P D_I P^{-1}$.

\subsection{The automatic hermiticity mechanism} \label{automatic}

In this subsection we give a brief review of 
the mechanism for suppressing the effect of $\hat{H}_{Qa}$ after a long time development 
of some state $| \psi(t) \rangle $, which obeys the Schr\"{o}dinger equation 
$i \hbar \frac{d}{dt} | \psi (t) \rangle = \hat{H} | \psi (t) \rangle $, 
by following refs.\cite{Nagao:2010xu, Nagao:2011za}. 
We introduce $| \psi'(t) \rangle $ by 
$| \psi' (t) \rangle = P^{-1} | \psi (t) \rangle$, and expand it as 
$| \psi'(t) \rangle = \sum_i a_i(t) | e_i \rangle $. 
Then $| \psi(t) \rangle $ can be written in an expanded form as 
$| \psi(t) \rangle = \sum_i a_i(t) | \lambda_i \rangle $. 
Since $ | \psi'(t) \rangle$ obeys 
$i \hbar \frac{d}{dt} | \psi'(t) \rangle =D | \psi'(t) \rangle$, 
the time development of $| \psi(t) \rangle$ from some time $t_0$ is calculated as 
\begin{eqnarray}
| \psi(t) \rangle &=& P e^{- \frac{i}{\hbar} D (t-t_0)} | \psi'(t_0) \rangle \nonumber\\
&=& \sum_i a_i(t_0) 
e^{ \frac{1}{\hbar} \left( \text{Im} \lambda_i - i \text{Re} \lambda_i \right) (t-t_0)}       
| \lambda_i \rangle . \
\end{eqnarray}

Now we assume that the anti-Hermitian part of $\hat{H}$ is bounded from above. 
We point out that this is a natural assumption in our CAT 
because this allows the whole FPI 
$=\int e^{\frac{i}{\hbar}S} {\cal D} \text{path}$ to converge. 
Indeed, this integral diverges unless the imaginary part of the action $S_I$ 
is bounded from below. 
In ref.\cite{Nagao:2011is}, to prevent the kinetic term from blowing up for 
$\dot{q}\rightarrow \pm \infty$, we gave a condition 
$\text{Im}(m) \geq 0$ on the mass $m$, 
which is equivalent to $\text{Im}\left(\frac{1}{m}\right) \leq 0$. 
In addition the imaginary part of the potential term $\text{Im}(V)$ 
has to be bounded from above. 
Thus the assumption of the boundedness of $H$ is needed 
to avoid the FPI $=\int e^{\frac{i}{\hbar}S} {\cal D} \text{path}$ 
being divergently meaningless. 

Based on this assumption 
we can crudely imagine that some of the $\text{Im} \lambda_i$ 
take the maximal value $B$. 
We denote the corresponding subset of $\{ i \}$ as $A$. 
Then, if a long time has passed, namely for large $t-t_0$, 
the states with $\text{Im} \lambda_i |_{i \in A}$ survive and contribute most in the sum. 
To show how $| \psi(t) \rangle $ is effectively described for large $t-t_0$, 
we introduce a diagonalized Hamiltonian $\tilde{D}_{R}$ as 
\begin{equation}
\langle e_i | \tilde{D}_{R} | e_j \rangle \equiv 
\left\{ 
 \begin{array}{cc}
      \langle e_i | D_R | e_j \rangle =\delta_{ij} \text{Re} \lambda_i  & \text{for} \quad i \in A , \\
      0 &\text{for} \quad i \not\in A , \\ 
 \end{array}
\right. \label{DRtilder}
\end{equation}
and define $\hat{H}_{\text{eff}}$ by 
$\hat{H}_{\text{eff}} \equiv P \tilde{D}_{R} P^{-1}$. 
Since $(\tilde{D}_{R})^{\dag} = \tilde{D}_{R}$, 
$\hat{H}_{\text{eff}}$ is $Q$-Hermitian, 
$\hat{H}_{\text{eff}} ^{\dag^Q} =\hat{H}_{\text{eff}}$, 
and satisfies $\hat{H}_{\text{eff}} | \lambda_i \rangle = \text{Re} \lambda_i | \lambda_i \rangle$. 
Furthermore, we introduce $| \tilde\psi(t) \rangle \equiv \sum_{i \in A}  a_i(t) | \lambda_i \rangle $. 
Then $| \psi(t) \rangle$ is approximately estimated as 
\begin{eqnarray}
| \psi(t) \rangle 
&\simeq& e^{ \frac{1}{\hbar} B (t-t_0)} 
\sum_{i \in A}  a_i(t_0) e^{-\frac{i}{\hbar} {\text Re} \lambda_i (t-t_0)} | \lambda_i \rangle \nonumber\\
&=&e^{ \frac{1}{\hbar} B (t-t_0)}  e^{-\frac{i}{\hbar} \hat{H}_{\text{eff}} (t-t_0)} 
| \tilde\psi(t_0) \rangle 
= | \tilde\psi(t) \rangle . \label{psiprimetket}
\end{eqnarray}
The factor $e^{ \frac{1}{\hbar} B (t-t_0)} $ in eq.(\ref{psiprimetket}) 
can be dropped out by normalization. 
Thus we have effectively obtained a $Q$-Hermitian Hamiltonian $\hat{H}_{\text{eff}}$ 
after a long time development.

Indeed the normalized state 
\begin{equation}
| \psi(t) \rangle_{N} 
\equiv \frac{1}{\sqrt{ \langle {\psi}(t) |_Q ~{\psi}(t) \rangle} } | {\psi}(t) \rangle 
\simeq \frac{1}{\sqrt{ \langle \tilde{\psi}(t) |_Q ~\tilde{\psi}(t) \rangle} } | \tilde{\psi}(t) \rangle 
\equiv | \tilde{\psi}(t) \rangle_{N}
\end{equation} 
originally obeys the slightly modified Schr\"{o}dinger equation,
\begin{equation}
i\hbar \frac{\partial}{ \partial t} | \psi(t) \rangle_{N} 
= \hat{H}_{Qh} | \psi(t) \rangle_{N} 
+ \left( \hat{H}_{Qa} -{}_{N} \langle \psi(t) |_Q \hat{H}_{Qa} | \psi(t) \rangle_{N} \right) | \psi(t) \rangle_{N} , \label{sch}
\end{equation} 
but after a long time it time-develops as 
$| \tilde{\psi}(t) \rangle_{N} =e^{-\frac{i}{\hbar} \hat{H}_{\text{eff}} (t-t_0)} | \tilde{\psi}(t_0) \rangle_{N}$, 
i.e. it obeys the Schr\"{o}dinger equation  
\begin{equation}
i\hbar \frac{\partial}{ \partial t} | \tilde\psi(t) \rangle_{N} = \hat{H}_{\text{eff}} | \tilde\psi(t) \rangle_{N}.
\end{equation}
We see that the time dependence of the normalization factor 
has disappeared.

On the other hand, we define the expectation value of some operator ${\cal O}$ by 
\begin{eqnarray}
\bar{\cal O}_Q(t) 
&\equiv&  {}_{N} \langle \psi(t) |_Q {\cal O} | \psi(t) \rangle_{N} 
=  {}_{N} \langle \psi(t_0) |_Q {\cal O}_{QH}(t-t_0) | \psi(t_0) \rangle_{N}  \nonumber \\
&\simeq&  {}_{N} \langle \tilde\psi(t) |_Q {\cal O} | \tilde\psi(t) 
\rangle_{N} 
= {}_{N} \langle \tilde\psi(t_0) |_Q \tilde{\cal O}_{QH}(t-t_0) | \tilde\psi(t_0) \rangle_{N}, \label{expOQAA}
\end{eqnarray} 
where we have introduced the time-dependent operator in the Heisenberg picture, 
\begin{eqnarray}
{\cal O}_{QH}(t-t_0) &\equiv& \frac{ \langle \psi(t_0) |_Q \psi(t_0) \rangle }{ \langle \psi(t) |_Q \psi(t) \rangle } 
e^{ \frac{i}{\hbar} \hat{H}^{\dag^Q} (t-t_0) } {\cal O} e^{ -\frac{i}{\hbar} \hat{H}(t-t_0) }  \nonumber \\
&\simeq&
e^{ \frac{i}{\hbar} \hat{H}_{\text{eff}} (t-t_0) } {\cal O}  
e^{ -\frac{i}{\hbar} \hat{H}_{\text{eff}}(t-t_0) } 
\equiv \tilde{\cal O}_{QH}(t-t_0) .
\end{eqnarray}
The time-dependent operator originally obeys the slightly modified Heisenberg equation, 
\begin{eqnarray}
&& i\hbar \frac{\partial}{ \partial t} {\cal O}_{QH}(t-t_0)  \nonumber \\
&=& [ {\cal O}_{QH}(t-t_0) , \hat{H}_{Qh} ] 
+ \left\{  {\cal O}_{QH}(t-t_0) , \hat{H}_{Qa} -  {}_{N} \langle \psi(t) |_Q \hat{H}_{Qa} | \psi(t) \rangle_{N} \right\} , \label{hei}
\end{eqnarray}
but after a long time development it obeys the Heisenberg equation 
\begin{equation}
i\hbar \frac{\partial}{ \partial t} \tilde{\cal O}_{QH}(t-t_0) 
= \frac{i}{\hbar} [ \hat{H}_{\text{eff}}, \tilde{\cal O}_{QH} (t-t_0)].
\end{equation}

As we have seen above, the non-Hermitian Hamiltonian $\hat{H}$ has automatically 
become a Hermitian one $\hat{H}_{\text{eff}}$ 
with the proper inner product $I_Q$ and a long time development.

\section{Nice properties of the expectation value $\langle {\cal O} \rangle^{BA} $}

In a future-included version of the CAT, 
$\langle {\cal O} \rangle^{BA}$, defined in eq.(\ref{expvalOBA0}), 
was considered as an expectation value of ${\cal O}$ in ref.\cite{Bled2006}, 
although this is a matrix element in the usual sense. 
A similar form was also considered in ref.\cite{AAV} in a different context. 
In this section we study this quantity further in the CAT 
and explicitly show that $\langle {\cal O} \rangle^{BA}$ has nice properties: it allows us to have 
the Heisenberg equation, Ehrenfest's theorem, and a conserved probability current density. 
These properties strongly suggest that $\langle {\cal O} \rangle^{BA} $ is a promising definition 
of the expectation value in the future-included CAT.

\subsection{Heisenberg equation}

From eqs.(\ref{schro_eq_Astate})(\ref{schro_eq_Bstate}), 
$| A(t) \rangle$ and $| B(t) \rangle$ time-develop as 
$| A(t) \rangle =\exp\left( -\frac{i}{\hbar} \hat{H} (t-T_A) \right) | A(T_A) \rangle$ and 
$| B(t) \rangle =\exp\left( -\frac{i}{\hbar} \hat{H}^\dag (t- T_B) \right) | B(T_B) \rangle$, 
where we have supposed a future final state $| B(T_B) \rangle$ and 
a past initial state $| A(T_A) \rangle$. 
As for eq.(\ref{expvalOBA0}), we note that the denominator $\langle B(t) | A(t) \rangle$ 
is constant in time, $\frac{d}{dt} \langle B(t) | A(t) \rangle=0$. 
We attempt to write the numerator 
as 
$\langle B(t) |  {\cal O}  | A(t) \rangle = \langle B(T_B) | \hat{O}_H(t) | A(T_A) \rangle$, 
where we have defined a Heisenberg operator 
\begin{equation}
\hat{O}_H(t) \equiv \exp\left( \frac{i}{\hbar} \hat{H} (t- T_B) \right) 
{\cal O}  \exp\left( -\frac{i}{\hbar} \hat{H} (t-T_A) \right)   
\end{equation}
obeying the Heisenberg equation 
$\frac{d}{dt}  \hat{O}_H(t) = \frac{i}{\hbar} [ H , \hat{O}_H(t) ]$. 
But we encounter  
\begin{eqnarray}
&&\hat{1}_H(t) = e^{\frac{i}{\hbar} \hat{H} (t_0-T_B) } , \label{1unusual} \\ 
&&\langle {\cal O}_1  {\cal O}_2 \rangle^{BA} 
= \frac{\langle B(T_B)| \hat{O}_{1H}(t) e^{\frac{i}{\hbar} \hat{H} (T_B-T_A) }
\hat{O}_{2H}(t)  | A(T_A) \rangle}
{ \langle B(T_B)| e^{-\frac{i}{\hbar} \hat{H} (T_B-T_A)}| A(T_A) \rangle }, 
\label{O1O2unusual}
\end{eqnarray}
which are not usual expressions. 
To avoid this situation we rewrite the numerator of eq.(\ref{expvalOBA0}) 
with some reference time $t_\text{ref}$, 
which can be chosen arbitrarily such that $T_A \le t_\text{ref} \le T_B$, 
as 
$\langle B(t) |  {\cal O}  | A(t) \rangle =
\langle B(t_\text{ref}) | \hat{O}_H(t ; t_\text{ref}) | A(t_\text{ref}) \rangle$, 
where we have defined another Heisenberg operator 
\begin{equation}
\hat{O}_H(t ; t_\text{ref}) \equiv \exp\left( \frac{i}{\hbar} \hat{H} (t- t_\text{ref}) \right) 
{\cal O}  \exp\left( -\frac{i}{\hbar} \hat{H} (t-t_\text{ref}) \right)  \label{OTAHeisenbergop}
\end{equation}
obeying the Heisenberg equation 
$\frac{d}{dt}  \hat{O}_H(t ; t_\text{ref})
= \frac{i}{\hbar} [ \hat{H} , \hat{O}_H(t ; t_\text{ref}) ]$.  
In contrast to eqs.(\ref{1unusual})(\ref{O1O2unusual}), 
we have the following usual expressions: 
\begin{eqnarray}
&&\hat{1}_H(t ; t_\text{ref}) = 1 ,  \\
&&\langle {\cal O}_1  {\cal O}_2 \rangle^{BA} 
= \frac{\langle B(t_\text{ref})| \hat{O}_{1H}(t ; t_\text{ref})
\hat{O}_{2H}(t ; t_\text{ref})  | A(t_\text{ref}) \rangle}
{ \langle B(t_\text{ref})| A(t_\text{ref}) \rangle } .
\end{eqnarray}
So we adopt the expression (\ref{OTAHeisenbergop}) for the Heisenberg operator 
in our theory.

Before finishing this subsection we make a remark on 
$\langle {\cal O} \rangle^{BA}$ and the contour $C$ in the path integral. 
The expectation value $\langle {\cal O} \rangle^{BA}$ is not real 
even for Hermitian ${\cal O}$, 
and would usually become so complicated that it would typically have values 
all over the complex plane ${\mathbf C}$.
This is in contrast to the expectation value $\langle {\cal O} \rangle^{AA}$ 
in the future-not-included theory, which is real for Hermitian ${\cal O}$. 
For both $\langle {\cal O} \rangle^{BA}$ and $\langle {\cal O} \rangle^{AA}$ 
there is no problem in deforming the integration contour $C$ at each time  
arbitrarily.

\subsection{Ehrenfest's theorem}

In this subsection we derive Ehrenfest's theorem. 
Utilizing the following relation 
\begin{eqnarray}
\frac{d}{dt} \langle {\cal O} \rangle^{BA} 
&=& \langle  \frac{i}{\hbar} [ \hat{H} , {\cal O} ]   \rangle^{BA} , 
\end{eqnarray} 
where we use $\hat{H}$ given in eq.(\ref{expHhat}), we obtain 
\begin{eqnarray}
&&\frac{d}{dt} \langle \hat{q}_{new} \rangle^{BA} 
= \frac{1}{m} \langle \hat{p}_{new} \rangle^{BA} ,  \label{dqcdt}  \\
&&\frac{d}{dt} \langle \hat{p}_{new} \rangle^{BA} 
= - \langle V'(\hat{q}_{new}) \rangle^{BA} .   \label{dpcdt} 
\end{eqnarray}
We note that eq.(\ref{dqcdt}) is consistent with the momentum relation 
derived via the path integral in ref.\cite{Nagao:2011is}. 
Substituting eq.(\ref{dqcdt}) for eq.(\ref{dpcdt}), we obtain Ehrenfest's theorem, 
\begin{equation}
m\frac{d^2}{dt^2} \langle \hat{q}_{new} \rangle^{BA} 
= - \langle V'(\hat{q}_{new}) \rangle^{BA} . \label{ehrenfest}
\end{equation}
We have thus checked that $\langle {\cal O} \rangle^{BA}$ 
provides the saddle point development with $t$.


\subsection{Conserved probability current density}

In this subsection we show that a conserved probability current density can be constructed 
in the future-included theory. 
First we define a probability density $\rho$ by 
\begin{equation}
\rho 
\equiv \frac{\psi_B(q)^{*_q} \psi_A(q)}{\langle B | A \rangle} 
=  \frac{ \langle B | q \rangle_{new}  ~{}_m\langle_{new}~ q | A \rangle }{ \langle B | A \rangle } ,
\end{equation}
where $\psi_A(q)$ and $\psi_B(q)$ are introduced 
in eqs.(\ref{psi_A(q)})(\ref{psi_B(q)}) respectively. 
This $\rho$ satisfies $\int_C dq \rho =1$, 
where $C$ is an arbitrary contour running from $-\infty$ to $\infty$ 
in the complex $q$-plane. 
Then defining a probability current density $j$ by 
\begin{equation}
j(q,t) =  \frac{ \frac{i\hbar}{2m} \left(  \frac{\partial \tilde{\psi}_B^{*_q}  }{\partial q}  
\psi_A - \tilde{\psi}_B^{*_q}   \frac{\partial \psi_A }{\partial q} \right) }{ \langle B | A \rangle } , 
\end{equation}
we have the continuity equation 
$\frac{\partial \rho}{\partial t } +  \frac{\partial }{\partial q} j(q,t) = 0$. 
Therefore, probability interpretation seems to work formally with this $\rho$, 
although $\rho$ is complex.

\section{Correspondence principle to ordinary quantum mechanics} \label{realobservables}
 

The future-included theory may look exotic because it includes time integration 
over not only the past but also the future. 
For such a theory to be viable it is very important to 
recapture the usual quantum mechanics even approximately from the future-included theory. 
Indeed, in ref.\cite{Bled2006}, 
such a possibility is speculated upon. 
We first examine the argument in ref.\cite{Bled2006}, and find that 
there are points to be improved. 
Next we study eq.(\ref{expvalOBA0}) carefully by utilizing the proper inner product and 
the mechanism of suppressing the anti-Hermitian part of the Hamiltonian, 
and propose the correspondence principle of the future-included theory 
to ordinary quantum mechanics.

\subsection{Former attempt to see the correspondence}

We briefly review the speculation to obtain the correspondence 
in ref.\cite{Bled2006} and see that there are points to be improved. 
In ref.\cite{Bled2006}, it was suggested that eq.(\ref{expvalOBA0}) is rewritten as 
\begin{equation}
\langle {\cal O} \rangle^{BA} 
=\frac{  \langle A(t) | B(t) \rangle\langle B(t) |  {\cal O}  | A(t) \rangle }
{ \langle A(t) | B(t) \rangle \langle B(t) | A(t) \rangle } , \label{expvalOBA2} 
\end{equation}
and an attempt was made to approximate $ | B(t) \rangle\langle B(t) |$ as 
\begin{equation}
| B(t) \rangle  \langle B(t) | 
\simeq \frac{1}{N} \sum_w | w \rangle  \langle w |  
= \frac{1}{N} 1  \label{BtketBtbrasimeq1}
\end{equation} 
for any $t$ except for times only slightly later than the early Big Bang time, 
where $N$ denotes the number of some 
orthonormal basis states $| w \rangle$ $(w=1, 2, \cdots , N)$ 
such that $\langle w | w' \rangle=\delta_{w, w'}$, 
with the assumption that the system is sufficiently ergodic. 
If we admit that eq.(\ref{BtketBtbrasimeq1}) is a good approximation, then 
$\langle  {\cal O} \rangle^{BA}$ becomes the expectation value in the future-not-included theory, 
\begin{equation}
\langle  {\cal O} \rangle^{BA} \simeq \langle  {\cal O} \rangle^{AA} ,
\end{equation}
where $\langle  {\cal O} \rangle^{AA}$ is given in eq.(\ref{OAA}). 
But, is eq.(\ref{BtketBtbrasimeq1}) really a good approximation?

Eq.(\ref{BtketBtbrasimeq1}) cannot be true at all $t$.  
In fact, using eq.(\ref{schro_eq_Bstate}) we obtain  
\begin{eqnarray}
\frac{d}{dt} \left( | B(t) \rangle \langle B(t) | \right) 
&=& -\frac{i}{\hbar}  [ \hat{H}_B^h , | B(t) \rangle \langle B(t) |  ] 
- \frac{i}{\hbar}  \{ \hat{H}_B^a , | B(t) \rangle \langle B(t) | \} \nonumber \\ 
&\simeq&
-2 \frac{i}{\hbar} \hat{H}_B^a ,  \label{ddtBketBbra}
\end{eqnarray} 
where $\hat{H}_B^h=  \frac{\hat{H}_B + \hat{H}_B^\dag}{2}$ and 
$\hat{H}_B^a=  \frac{\hat{H}_B - \hat{H}_B^\dag}{2}$, 
and in the second equality 
we have used eq.(\ref{BtketBtbrasimeq1}). 
Thus we have 
$| B(t) \rangle \langle B(t) | \simeq C_1 \exp\left[-2 \frac{i}{\hbar} \hat{H}_B^a t \right]$, 
where $C_1$ is some constant.   
If we choose $C_1$ so that we have 
$| B(t) \rangle \langle B(t) | 
\simeq \frac{1}{N} \exp\left[-2 \frac{i}{\hbar} (t - T_B ) \hat{H}_B^a \right]$, 
then for $t= T_B$ this becomes 
$| B( T_B) \rangle \langle B( T_B) | \simeq \frac{1}{N} 1$. 
So eq.(\ref{BtketBtbrasimeq1}) becomes a good approximation near the far future time $T_B$, 
but it is not so good for general time $t$. 
%

\subsection{Our analysis of $\langle {\cal O} \rangle^{BA}$}

We analyze eq.(\ref{expvalOBA2}) more carefully by utilizing the proper inner product 
and the automatic hermiticity mechanism. 
Expanding $| B(T_B) \rangle$ as 
$| B(T_B) \rangle = \sum_i b_i | \lambda_i \rangle_B$, we obtain 
\begin{eqnarray}
| B(t) \rangle \langle B(t) | 
&=&  
e^{-i \hat{H}_B (t - T_B)} | B(T_B) \rangle  
\langle B(T_B) |_{Q_B}  e^{i \hat{H}_B^{\dag^{Q_B}} (t - T_B) }  
Q_B^{-1} \nonumber \\ 
&=&
\sum_{i,j} b_i  b_j^* e^{i  \text{Re}(\lambda_j - \lambda_i) (t - T_B)} 
e^{ \text{Im}(\lambda_j + \lambda_i) (T_B - t)} 
| \lambda_i \rangle_B  ~{}_B \langle \lambda_j |  \nonumber \\ 
&\simeq&  
\frac{\int_{t-\Delta t}^{t+ \Delta t} | B(t) \rangle \langle B(t) | dt }{ \int_{t-\Delta t}^{t+ \Delta t} dt } \nonumber \\
&\simeq&  
\sum_i | b_i |^2  e^{ 2\text{Im}(\lambda_i) (T_B - t)} | \lambda_i \rangle_B  ~{}_B \langle \lambda_i |  
\nonumber \\
&\simeq& e^{ 2B (T_B - t)}  Q_2  \quad  \text{for large $T_B - t$} , 
\label{Q2_0}
\end{eqnarray}
where $Q_B$ and $\hat{H}_B^{\dag^{Q_B}}$ are given 
in eqs.(\ref{QB_rel_Q-1})(\ref{HBdagQB}) respectively. 
In the third line we have smeared the present time $t$ a little bit, 
and then, since the off-diagonal elements wash to $0$, we are led to the fourth line. 
In the last line we have used the automatic hermiticity mechanism  
for large $T_B - t$, and $Q_2$ is given by 
\begin{eqnarray}
Q_2
&=&\sum_{i \in A} | b_i |^2 | \lambda_i \rangle_B  ~{}_B \langle \lambda_i |  \nonumber \\ 
&=&
\sum_{i \in A} G(\hat{H}_{\text{eff}} + i B \Lambda_A )^\dag | \lambda_i \rangle_B 
~{}_B \langle \lambda_i | 
G(\hat{H}_{\text{eff}} + iB \Lambda_A )
\nonumber \\ 
&=&
\tilde{G}(\hat{H}_{\text{eff}} )^\dag Q \Lambda_A \tilde{G}(\hat{H}_{\text{eff}} ), \label{newQ2}
\end{eqnarray}
where, in the second and third lines, 
supposing that $\text{Re} \lambda_i$'s are not degenerate, 
we have introduced 
$\Lambda_A \equiv \sum_{i \in A} | \lambda_i \rangle \langle \lambda_i |_Q$, 
and 
functions $G$ and $\tilde{G}$ such that 
$G(\text{Re} \lambda_i + iB)= \tilde{G}(\text{Re} \lambda_i ) = b_i$, 
and 
we have used $| \lambda_i \rangle_B = Q | \lambda_i \rangle$, and 
${}_B\langle \lambda_i | G(\text{Re} \lambda_i + iB)   =  {}_B\langle \lambda_i | G(\hat{H}_{\text{eff}} + iB \Lambda_A ) 
~\text{for} ~i \in A$.  
We note that 
$Q \Lambda_A=Q \sum_{i \in A} | \lambda_i \rangle \langle \lambda_i |_Q$ is Hermitian, and so is $Q_2$.

Next we expand $| A(t) \rangle$ as 
$| A(t) \rangle \equiv \sum_i  a_i(t) | \lambda_i \rangle $, 
and define 
$| \tilde A(t) \rangle \equiv \sum_{i \in A}  a_i(t) | \lambda_i \rangle $.  
Then for large $t-T_A$, 
since we have $| A(t) \rangle \simeq | \tilde A(t) \rangle$ 
by using the automatic hermiticity mechanism  
as in eq.(\ref{psiprimetket}),  
we can express eq.(\ref{expvalOBA2}) as 
\begin{equation}
\langle {\cal O} \rangle^{BA} 
\simeq
\frac{  \langle \tilde{A} (t) |_{Q_2}  {\cal O}  | \tilde{A}(t) \rangle }
{ \langle \tilde{A}(t) |_{Q_2}  \tilde{A}(t) \rangle }  
\quad \text{for large $T_B - t$ and large $t-T_A$}, \label{result_o_BA}
\end{equation}
where $Q_2$ is given by eq.(\ref{newQ2}). 
In eq.(\ref{result_o_BA}) $|\tilde{A}(t) \rangle$ is really the state of our whole universe 
as obtained from the initial state. 
From a classical point of view it is likely to be 
a superposition of many wildly different states representing 
narrow wave packets. 
In practice, since we live inside this universe, we come to know features 
that in the CAT 
are determined from $\langle B(t) |$, i.e. ``the future", 
rather than only from $| \tilde{A} (t) \rangle$. 
Information about such features coming from the future may partly stand 
in our memories and we can combine this information with 
information on $| \tilde{A} (t) \rangle$ to obtain a better and 
in some way more realistic replacement for $| \tilde{A} (t) \rangle$ 
which we call $|\tilde{\psi}_{memory}(t) \rangle$. 
We hope to return to this improvement of $| \tilde{A} (t) \rangle$ to $|\tilde{\psi}_{memory}(t) \rangle$ 
in a later article, but since this replacement plays effectively no important role in the 
present article, 
we shall just keep the notation  $| \tilde{A} (t) \rangle$ and bear in mind that it would be more realistic 
to call it $|\tilde{\psi}_{memory}(t) \rangle$.

\subsection{Our proposal of the correspondence} 

In ref.\cite{Bled2006} an expectation value in a future-not-included theory is 
defined by eq.(\ref{OAA}), 
but in refs.\cite{Nagao:2010xu, Nagao:2011za} and in eq.(\ref{expOQAA}) 
we defined a slightly different one by 
\begin{eqnarray}
\langle {\cal O} \rangle_Q^{AA} 
&\equiv&
\frac{  \langle A(t) |_Q {\cal O}  | A(t) \rangle }
{ \langle A(t) |_Q A(t) \rangle }. \label{expOQAA2} 
\end{eqnarray}
In this subsection we show that it is the latter definition of an expectation value 
that we obtain from a future-included theory. 
We first define $Q'$ by 
$Q' \equiv G(\hat{H})^\dag Q G(\hat{H}) =({P_{G^{-1}} }^\dag )^{-1} {P_{G^{-1}} }^{-1}$, 
where $P_{G^{-1}} \equiv G(\hat{H})^{-1} P$ diagonalizes $\hat{H}$: 
$(P_{G^{-1}})^{-1} \hat{H} P_{G^{-1}} = P^{-1} \hat{H} P = D$.  
In addition, we introduce 
$| \lambda_i \rangle^{G^{-1}} \equiv G(\hat{H})^{-1} | \lambda_i \rangle$, 
so that $| \lambda_i \rangle^{G^{-1}}$ is $Q'$-orthogonal, i.e., orthogonal with regard to 
the proper inner product $I_{Q'}$: 
$I_{Q'} (   | \lambda_i \rangle^{G^{-1}} ,  | \lambda_j \rangle^{G^{-1}} ) 
\equiv {}^{G^{-1}}\langle \lambda_i | Q' | \lambda_j \rangle^{G^{-1}} 
=\delta_{ij}$. 
We use the automatic Hermiticity mechanism for large $t-T_A$. 
Then, since $| A(t) \rangle$ behaves as 
$| \tilde A(t) \rangle \equiv \sum_{i \in A}  a_i(t) | \lambda_i \rangle$, 
$Q'$ used in the normalized matrix element 
$\langle {\cal O} \rangle_{Q'}^{AA}$ is estimated in the subspace restricted by $A$ 
as follows: 
\begin{eqnarray}
Q' 
&\simeq&  G(\hat{H}_{\text{eff}} + iB \Lambda_A )^\dag Q \Lambda_A G(\hat{H}_{\text{eff}} + iB \Lambda_A ) 
\quad {\text{for the restricted subspace}}  \nonumber \\
&=&
\tilde{G}(\hat{H}_{\text{eff}} )^\dag Q \Lambda_A \tilde{G}(\hat{H}_{\text{eff}} ) \nonumber \\
&=& Q_2 ,  
\end{eqnarray}
where in the last equality we have used eq.(\ref{newQ2}).  
Then with the inner product $I_{Q'}$ 
the expectation value in a future-not-included theory is expressed as 
\begin{eqnarray}
\langle {\cal O} \rangle_{Q'}^{AA} 
&=&
\frac{  \langle A(t) |_{Q'} {\cal O}  | A(t) \rangle }
{ \langle A(t) |_{Q'} A(t) \rangle } \nonumber \\
&\simeq&  
\frac{  \langle \tilde{A}(t) |_{Q_2} {\cal O}  | \tilde{A}(t) \rangle }
{ \langle \tilde{A}(t) |_{Q_2} \tilde{A}(t) \rangle } 
\quad \text{for large $t-T_A$} .   
\label{expvalinfuturenotwithlongtime} 
\end{eqnarray}
Comparing eq.(\ref{result_o_BA}) with eq.(\ref{expvalinfuturenotwithlongtime}), we obtain 
the following correspondence: 
\begin{equation}
\langle {\cal O} \rangle^{B A} ~\text{for large $T_B-t$ and large $t-T_A$}
\quad \simeq   \quad 
\langle {\cal O} \rangle_{Q'}^{AA} ~\text{for large $t-T_A$}. \label{correspondence}
\end{equation}
This relation means that 
the future-included theory for large $T_B-t$ and large $t-T_A$ is almost 
equivalent to the future-not-included theory with the proper inner product 
for large $t-T_A$. 
This equivalence suggests that 
the future-included theory is {\em not excluded}, although it seems exotic.

\subsection{A seemingly time reversal symmetry problem}

Finally, let us discuss a seemingly time reversal symmetry problem. 
In the functional integral formulation with future included there seems 
to be no difference between the past and future time directions. 
In the light of this, it seems strange that we obtain a description
in terms of the form $\langle {\cal O} \rangle_{Q'}^{AA}$ rather than the form $\langle {\cal O} \rangle_{Q''}^{BB}$ with another appropriate operator $Q''$ 
which means that we use the future instead of the past. 
Essentially $\langle {\cal O} \rangle^{BA}$ can be rewritten as either 
$\langle {\cal O} \rangle_{Q'}^{AA}$ or $\langle {\cal O} \rangle_{Q''}^{BB}$, 
as we like. However, we know phenomenologically that the past state
influences the present state, while the future state does not influence the
present state so much,  so we can choose the expression 
$\langle {\cal O} \rangle_{Q'}^{AA}$. 
But we hope that we can explain why it is chosen. 
Our present universe is at a low temperature and has a low energy density, 
while the imaginary part of the action is very small. 
On the other hand, the situation of an early universe -- high temperature and high energy density -- 
is very different from our present era. 
So we have some possibility that 
in the early universe there was a period in which the imaginary part of the Lagrangian was 
much more important than later or in the future. 
This possibility is speculated on in refs.\cite{Bled2006,own,Nielsen:2007mj}. 
It is an open problem to show this explicitly, but 
if it is proven to be true, then 
the reason that 
the present expectation value is to be described approximately in terms of a model 
with an initial state determined picture meaning $\langle {\cal O} \rangle_{Q'}^{AA}$ 
would be that the solution is mainly determined by a requirement 
involving the imaginary part of the Lagrangian 
in the early universe, while at most small clear signals come from the future. 
Thus $\langle {\cal O} \rangle^{BA}$ should be rewritten as $\langle {\cal O} \rangle_{Q'}^{AA}$ rather than $\langle {\cal O} \rangle_{Q''}^{BB}$. 
We speculate that 
the past and the future are physically different in the sense that the past is $S_I$-dominant, 
while the future is not $S_I$-dominant, which essentially causes 
the time reversal symmetry breaking.

\section{Summary and outlook}

We have studied a future-included version of a complex action theory (CAT), 
which includes time integration over not only past but also future. 
In ref.\cite{Bled2006} a correspondence of the theory to 
ordinary quantum mechanics, i.e. a future-not-included version of a real action theory (RAT), 
was speculated upon. 
In this paper, studying the quantity 
$\langle {\cal O} \rangle^{BA}$ defined in eq.(\ref{expvalOBA0}) 
more carefully and in detail 
using both the automatic hermiticity mechanism\cite{Nagao:2010xu,Nagao:2011za} 
and the complex coordinate formalism\cite{Nagao:2011za}, 
we have confirmed that, even if future is fundamentally included 
in the formalism, it leads to only minute deviations 
from ordinary quantum mechanics. 
This correspondence principle is one of the main results obtained in this paper.

In section 2 we reviewed our complex coordinate formalism and gave a theorem for matrix elements of the form 
${}_m\langle_{new}~ q' ~\text{or}~ p'|
{\cal O}( \hat{q}_{new}, \hat{q}_{new}^\dag, \hat{p}_{new}, \hat{p}_{new}^\dag) | q'' ~\text{or}~ p'' \rangle_{new}$, 
which states that we can ignore the 
anti-Hermitian terms in $\hat{q}_{new}$, $\hat{q}_{new}^\dag$, 
$\hat{p}_{new}$ and $\hat{p}_{new}^\dag$ 
provided that we are satisfied with the result in the distribution sense. 
In section 3, after explaining the two states $\langle B(t) |$ and $| A(t) \rangle$ 
definable from their respective functional integrals over future and past 
following ref.\cite{Bled2006}, 
we gave the two slightly improved wave functions 
$\psi_A(q) = {}_m\langle_{new}~ q | A(t) \rangle$ and 
$\psi_B(q) = {}_m\langle_{new}~ q | B(t) \rangle$ 
based on the complex coordinate formalism\cite{Nagao:2011za}, 
so that they are properly defined even for complex $q$. 
Then, reviewing the proper inner product for the Hamiltonian $\hat{H}$ in the future-not-inclulded theory, 
we constructed the proper inner product for the other Hamiltonian $\hat{H}_B$ 
in the future-included theory. 
We also reviewed the automatic hermiticity mechanism 
for $\hat{H}$ in the future-not-included theory. 
In section 4 we studied the behavior of $\langle {\cal O} \rangle^{BA}$. 
We showed that it allows us to have the Heisenberg equation and Ehrenfest's theorem. 
We also obtained the momentum relation 
$ \langle \hat{p}_{new} \rangle^{BA} = m \frac{d}{dt} \langle \hat{q}_{new} \rangle^{BA}$ 
in eq.(\ref{dqcdt}), which is consistent with 
the result of ref.\cite{Nagao:2011is}. 
Furthermore, we constructed a conserved probability current density by 
using both $|A \rangle$ and $|B \rangle$. 
We have thus checked that $\langle {\cal O} \rangle^{BA}$ 
provides the saddle point development with $t$ 
and works as an expectation value though its appearance is a matrix element.

In section 5 we analyzed the quantity $\langle {\cal O} \rangle^{BA}$, 
and derived the correspondence of the future-included theory to ordinary quantum mechanics. 
Showing that usual physics is approximately obtained from the future-included theory is 
very important, 
because the future-included theory seems excluded phenomenologically 
from two unusual points: 
the existence of the imaginary part of the action $S$ and that of the future state. 
We first reviewed the speculation on the correspondence in ref.\cite{Bled2006}, and 
made it clear that there are points to be improved in the argument. 
Then we studied $\langle {\cal O} \rangle^{BA}$ with more care concerning 
the inner product being obtained 
by using both the complex coordinate formalism and the automatic hermiticity mechanism, 
and showed that the quantity $\langle {\cal O} \rangle^{BA}$ becomes 
an expectation value with a different inner product defined in a future-not-included theory. 
Next we showed that the inner product can be interpreted 
as one of the possible proper inner products realized in the future-not-included theory. 
Thus we have obtained the correspondence principle in eq.(\ref{correspondence}), i.e. 
$\langle {\cal O} \rangle^{B A} ~\text{for large $t-T_A$ and  large $T_B-t$}
\quad \simeq   \quad 
\langle {\cal O} \rangle_{Q'}^{AA} ~\text{for large $t-T_A$}$, 
where $T_A$, $T_B$, and $t$ are the past initial time, the future final time, 
and the present time, 
respectively, and $Q'$ is a Hermitian operator used to define the proper inner product. 
This relation means that the future-included theory for large $T_B-t$ and 
large $t-T_A$ is almost 
equivalent to the future-not-included theory with the proper inner product 
for large $t-T_A$.

Thus it is not excluded that fundamentally the action is complex. 
Indeed, the reality of an action in ordinary quantum theory 
can be regarded as a restriction on parameters in the action, 
so it is a benefit of our theory 
that we can have a more general action by getting rid of the restriction. 
Also, since we found that the effects of backward causation are in practice small, 
a theory with a functional integral of future time is not excluded and 
what happens in the future can actually in principle act back on us today. 
In addition, as we have seen in this paper, the future-included theory 
looks more elegant in the functional integral formulation and 
shows more cleanly the saddle points providing the classical 
approximation than the future-not-included theory 
that we studied previously 
in refs.\cite{Nagao:2010xu,Nagao:2011za}, where we encountered 
additional complicated terms. 
Furthermore, the future-included theory 
can yield the future-not-included theory with the proper inner product. 
These are the advantages of the future-included theory.

As our next steps, what should we study?
First we note that in 
the above correspondence we have the Hermitian operator $Q'$. 
It is a priori non-local, but phenomenologically it should be local. 
We expect that it becomes effectively local 
somehow in some reasonable approximation. 
It would be desirable to invent some mechanism for making it effectively local. 
Also, it is very important to study the dynamics in some concrete model of the future-included theory. 
Indeed, since we already have the complex coordinate formalism and so on, it would be possible 
to perform the analyses. 
Furthermore, a transactional interpretation\cite{Cramer} of quantum mechanics 
is discussed in refs.\cite{Vaxjo2009}\cite{newer2} based on the future-included theory. 
It would be interesting to study in detail the relation between the interpretation and 
the future-included theory. 
We hope to report studies on these problems in the future.


\section*{Acknowledgements}

The work of K.N. was supported in part by 
Grant-in-Aid for Scientific Research (Nos. 18740127 and 21740157) 
from the Ministry of Education, Culture, Sports, Science and Technology (MEXT, Japan). 
K.N. would like to thank the members and visitors of NBI for their kind hospitality, 
and T.~Asakawa and I.~Tsutsui for useful discussions. 
H.B.N. is grateful to NBI for allowing him to work at the institute as emeritus.

\appendix

\section{Derivation of $\hat{H}_B$}\label{der_H_B}

The Feynman path integral (FPI) in the complex action theory (CAT) is described with 
the following Lagrangian, with a single degree of freedom for simplicity: 
\begin{equation}
L(q(t), \dot{q}(t))=\frac{1}{2}m \dot{q}^2- V(q) , \label{lagrangian}
\end{equation}
where $V(q)$ is a potential term defined by 
$V(q)=\sum_{n=2}^\infty b_n q^n$. 
This Lagrangian has the same form as that in the real action theory (RAT), but 
since we consider it in the CAT, 
$m$, $q$ and any other parameters included in $V(q)$ are complex in general. 
We consider the integrand of the FPI 
$\exp\left( \frac{i}{\hbar} \int L dt \right)$ 
by discretizing the time direction and writing $\dot{q}$ as 
$\dot{q} = \frac{ q(t+ dt) - q(t) }{dt}$, 
where $dt$ is assumed to be a small quantity. 
Since we use the Schr\"{o}dinger representation for wave functions, 
to avoid confusion with the Heisenberg representation 
we introduce the notations $q_t \equiv q(t)$ and  
$q_{t+dt} \equiv q(t+dt)$, which we regard as independent variables.

In ref.\cite{Nagao:2011is} we explicitly examined the momentum 
and the Hamiltonian since it is not trivial whether we can use the same forms as those 
in the RAT, 
because they includes a quantity at time $t+dt$, $q_{t+dt}$, 
which is somehow unclear from the point of view
of quantum mechanics unless 
we define it properly including the fluctuation in the time development from 
a quantity at time $t$. 
Thus we are motivated to examine 
them by describing $q_{t+dt}$ properly via FPI.  
We briefly explain how we derived $\hat{H}$ 
in ref.\cite{Nagao:2011is}.

In FPI, the time development of some wave function 
${}_m \langle_{new}~ q_t | \psi(t) \rangle$ at time $t$ to $t+dt$ is described by  
\begin{equation}
{}_m \langle_{new}~ q_{t+dt} | \psi(t+dt) \rangle 
= \frac{1}{\alpha(dt)} \int_C e^{\frac{i}{\hbar} dt L(q,\dot{q})}
{}_m \langle_{new}~ q_{t} | \psi(t) \rangle d q_t , \label{time_dev_qbraAket}
\end{equation}
where $L(q, \dot{q})$ is given by eq.(\ref{lagrangian}), and 
$C$ is an arbitrary path running from $-\infty$ to $\infty$ in the complex plane. 
In addition, $\alpha(dt)$ is a $dt$-dependent normalization factor, 
which is properly fixed later. 
In ref.\cite{Nagao:2011is}, to derive the momentum relation 
$p=\frac{\partial L}{\partial \dot{q}}$, 
we considered some wave function ${}_m \langle_{new}~ q_t | \xi \rangle$ that obeys 
\begin{eqnarray}
{}_m \langle_{new}~ q_t | \hat{p}_{new} | \xi \rangle 
&=&
\frac{\hbar}{i} \frac{\partial}{\partial q_t}~{}_m \langle_{new}~ q_t | \xi \rangle \nonumber \\ 
&=&
\frac{\partial L}{\partial \dot{q}}\left( q_t, \frac{\xi-q_t}{dt} \right) 
{}_m \langle_{new}~ q_t | \xi \rangle , \label{phat_p_xi2}
\end{eqnarray}
where $\xi$ is any number. 
Since the set $\left\{ |\xi \rangle \right\}$ is an approximately reasonable basis 
that has roughly completeness 
$1 \simeq \int_C d\xi | \xi \rangle ~{}_m \langle \text{anti} ~\xi |$ 
and orthogonality 
${}_m \langle \text{anti} ~\xi | \xi' \rangle \simeq \delta_c(\xi -\xi')$, 
where ${}_m \langle \text{anti} ~\xi |$ is a dual basis of $| \xi \rangle$, 
we can expand the wave function ${}_m \langle_{new}~ q_t | \psi(t) \rangle$ into 
a linear combination of ${}_m \langle_{new}~ q_t | \xi \rangle$ as 
${}_m \langle_{new}~ q_t | \psi(t) \rangle 
= \int_C d\xi  ~{}_m \langle_{new}~ q_t | \xi \rangle ~{}_m \langle \text{anti} ~\xi | \psi(t) \rangle$. 
Then, solving eq.(\ref{phat_p_xi2}), we can estimate the right-hand side 
of eq.(\ref{time_dev_qbraAket}) 
explicitly as follows: 
\begin{eqnarray}
{}_m \langle_{new}~ q_{t+dt} | \psi(t+dt) \rangle 
&=&
\frac{1}{\alpha(dt)} \int_{C'} d\xi
\int_{C} d q_t e^{\frac{i}{\hbar} dt L(q,\dot{q})}
~{}_m \langle_{new}~ q_t | \xi \rangle ~{}_m \langle \text{anti} ~\xi | \psi(t) \rangle
\nonumber \\
&\simeq& 
~{}_m \langle_{new}~  q_{t+dt} |  \exp \left( -\frac{i}{\hbar} \hat{H} dt 
\right)  |\psi(t) \rangle , \label{time_dev_qbraAket_k=0_3}
\end{eqnarray}
where $\hat{H}$ is given by eq.(\ref{expHhat}). 
Here we have taken $\alpha(dt)= \sqrt{\frac{2\pi i \hbar dt}{m}}$ so that 
both sides  of eq.(\ref{time_dev_qbraAket_k=0_3}) correspond to each other in the vanishing limit of $dt$. 
Then eq.(\ref{time_dev_qbraAket_k=0_3}) is reduced to 
$|\psi(t+dt) \rangle = e^{-\frac{i}{\hbar} \hat{H} dt}|\psi(t) \rangle$. 
Thus we have derived the Schr\"{o}dinger equation and found that 
the Hamiltonian $\hat{H}$ has the same form as that in the RAT 
starting from eq.(\ref{time_dev_qbraAket}). 
Such a derivation of the Schr\"{o}dinger equation is well known in the RAT~\cite{FPIbook}.

We can obtain the expression of $\hat{H_B}$ analogously to the calculation in ref.\cite{Nagao:2011is} 
just by noticing the following points. 
Performing the $*_q$ operation on eq.(\ref{time_dev_qbraAket}) we obtain 
$\langle \psi(t+dt)  |  q_{t+dt}  \rangle_{new}~ 
= \frac{1}{\alpha(dt)^*}
\int_C e^{-\frac{i}{\hbar} dt L(q,\dot{q})^{*_q} } 
\langle \psi(t) |  q_{t}  \rangle_{new}~ d q_t$. 
Defining $dt'=-dt$, we rewrite this as 
\begin{equation}
\langle \psi(t - dt')  |  q_{t - dt'} \rangle_{new}~ 
= \frac{1}{\alpha(-dt')^*} \int_C e^{ \frac{i}{\hbar} dt' L(q,\dot{q})^{*_q}  }
\langle \psi(t) |  q_{t}  \rangle_{new}~ d q_t , \label{bra_psi_t-deltat}
\end{equation}
where 
$L(q,\dot{q})^{*_q} = \frac{1}{2} m^* \dot{q}^2 - V(q)^{*_q}$ and 
$V(q)^{*_q}=\sum_{n= 2} b_n^* q^n$.  
On the other hand, the time development of the wave function 
$\langle B(t) | q_t \rangle$ at time $t$ to time $t-dt$ is described by  
\begin{equation}
\langle B(t-dt) | q'_{t- dt} \rangle_{new}~ =    
\frac{1}{\alpha(-dt)}
\int_{\text{path}(t-dt) = 
q'_{ t-dt }}  \langle B(t) | q_t \rangle_{new}~ e^{\frac{i}{\hbar} S_{t-dt ~\text{to} ~t}} D\text{path} . \label{pi_Bbra_deltat}
\end{equation}
Comparing this expression with eq.(\ref{bra_psi_t-deltat}), 
we can derive $\hat{H}_B$ in a similar way 
to the derivation of $\hat{H}$ in ref.\cite{Nagao:2011is}. 
Indeed, we obtain the Schr\"{o}dinger equation 
$|B (t-dt) \rangle = e^{\frac{i}{\hbar} \hat{H}_B dt}|B (t) \rangle$, 
where $\hat{H}_B$ is given just by the replacement of the coupling parameters 
and operators with their complex or Hermitian conjugates 
in the expression of $\hat{H}$, 
$\hat{H}_B  = \frac{1}{2m^*} (\hat{p}_{new}^{\dag})^2  + \sum_{n=2} b_n^* ( \hat{q}_{new}^{\dag})^n 
= \hat{H}^{\dag}$. 
Thus we have derived eq.(\ref{defHB}).


\end{document}